\pdfoutput=1
\documentclass[11pt]{article}

\usepackage[final]{acl}
\usepackage{times}
\usepackage{latexsym}
\usepackage{graphicx}
\usepackage{subcaption}
\usepackage{algorithm}
\usepackage{amsmath}
\usepackage{algpseudocode}
\usepackage{amsfonts}
\usepackage{amssymb}
\usepackage{todonotes} % for visible notes
\usepackage{enumitem}
\usepackage{booktabs}
\usepackage[T1]{fontenc}
\usepackage[utf8]{inputenc}

\usepackage{microtype}

\usepackage{inconsolata}

\title{Linking Scalar-Intensity Language to Structural Polarization with Validated Signed-Network Measures}

\author{
Zhijin Guo$^{1}$ \quad
Li Zhang$^{2}$ \quad
Tyler Bonnet$^{1,3}$ \quad
Janet B. Pierrehumbert$^{1}$ \quad
Xiaowen Dong$^{1}$ \\
$^{1}$University of Oxford \quad
$^{2}$University College London \quad
$^{3}$Imperial College London \\
\texttt{zhijin.guo@eng.ox.ac.uk}
}

\begin{document}
\maketitle

\begin{abstract}
Polarization in online communities is often studied through either language or interaction structure, but the two views are rarely connected within a unified framework. Prior work has linked them by constructing interaction graphs from human judgements of agreement and disagreement, leaving a gap between language as observed text and structure as an engineered representation of that text. We address this gap with a language-grounded signed-network pipeline that derives signed relations directly from conversational exchanges and links window-level language patterns to structural polarization over time. Before examining this relationship, we compare spectral and frustration-based polarization measures on synthetic benchmarks and real interaction networks. We find that frustration-based measures normalized by the graph's cycle-space capacity provide a more suitable basis for comparing polarization across networks of different sizes and densities. We therefore carry forward two complementary frustration-based measures: a weighted form that incorporates stance-model confidence and a count form based on edge signs. The weighted measure provides the better-behaved structural estimate and aligns more closely with polarization measured from human-labelled interactions, while the count measure reveals a stronger relationship with language. Across monthly Reddit Brexit discussions, greater prevalence of scalar-intensity language is associated with greater structural polarization, with a similar rank-level pattern in the human-labelled network. We find little evidence that language in one month predicts polarization in the next, whereas contemporaneous scalar-intensity prevalence provides useful information about polarization within the same month. Overall, our results demonstrate how language and interaction structure can be connected through signed networks while showing that the choice of structural polarization measure shapes the relationship that is observed.
\end{abstract}

\section{Introduction}

Polarization in online communities matters because it reshapes collective behaviour: it hardens group boundaries, encourages echo chambers, and can accelerate the spread of low-quality or misleading information \citep{bakshy2015exposure, cinelli2021echo, vosoughi2018spread}. Recent work also shows that polarization is not only structural but can be affective and tightly coupled to information diffusion dynamics \citep{lerman2024affective}. Yet empirical studies often split along a methodological divide: language-based analyses focus on what people say, while network-based analyses focus on who interacts with whom. By contrast, constructing \emph{signed} interaction graphs in which edge signs reflect (dis)agreement requires additional discourse interpretation and has only recently been supported at scale by resources such as DEBAGREEMENT \citep{pougue2021debagreement}. Accordingly, language-grounded signed-network polarization analysis remains a comparatively underexplored direction.

This creates a fundamental mismatch between how discourse is observed and how structure is modelled \citep{durrheim2025polarization}. Conversation logs provide text, whereas structural accounts of polarization are framed in terms of signed relations that encode agreement and disagreement between users, rooted in structural balance theory \citep{cartwright1956structural} and modern signed-network modelling \citep{huang2022pole}. In practice, however, links are rarely derived directly and systematically from the language they are meant to represent: network pipelines often ignore sign by treating edges as unsigned or collapsing interactions into binary links, while Natural Language Processing (NLP) pipelines commonly reduce discourse to sentiment, topics, or text similarity.

Two methodological issues follow from this gap. First, when signed relations are inferred from language, reducing them to hard labels discards information about model confidence and stance strength. Retaining continuous signed weights allows us to ask whether the magnitude of agreement and disagreement provides structural information beyond edge direction alone. Second, structural polarization scores are often compared across networks or time windows without first establishing that changes in the score reflect signed structure rather than differences in graph size or density. Without this validation, an apparent relationship between discourse and polarization may instead reflect changes in the amount or pattern of interaction.

In this paper, we examine how structural polarization in signed interaction networks relates to language use. We introduce a language-grounded signed-network pipeline that converts conversational exchanges into confidence-weighted agreement and disagreement ties and measures their structural polarization over time. This design lets us use language in two complementary ways: first, as the basis for constructing the signed network itself, and second, as a set of interpretable window-level signals that can be related to the resulting structural polarization series. In a case study of Reddit Brexit discussions, we examine scalar-intensity language, toxicity, and perplexity, and ask how these discourse signals relate to structural polarization contemporaneously and across time.

Before conducting the language analysis, we examine which structural polarization measures are suitable for comparison across changing interaction networks. We compare spectral Eigen-Sign approaches \citep{bonchi2019discovering} and frustration-based approaches \citep{doreian2009partitioning} on synthetic networks with planted two-camp structure and on the real monthly interaction graphs (Section~\ref{sec:benchmark_validation}). While both families can recover planted structure, their scores differ in their sensitivity to graph size and density. The cycle-normalized frustration measures provide the more stable basis for cross-window comparison, so we carry them into the language analysis. We use a weighted form that incorporates stance-model confidence as the primary structural measure, together with a count form based only on edge signs as a complementary view.

Using these measures, we find that the prevalence of scalar-intensity language is positively associated with the count-based frustration index, with a similar rank-level pattern when the network is constructed directly from human agreement and disagreement labels. The weighted frustration index provides a more stable measure of structure and aligns more closely with the human-labelled structural series, but shows a weaker language relationship, indicating that the two indices capture complementary aspects of the same signed networks. We further find little evidence that the previous month's language forecasts subsequent polarization at monthly resolution; instead, scalar-intensity prevalence carries useful contemporaneous information about structural polarization. Together, these results connect language and interaction structure while showing that the structural measure itself must be validated before discourse--polarization relationships are interpreted.

\noindent\textbf{Contributions.} The main contributions of our paper are as follows:
\begin{itemize}[leftmargin=*]

    \item \textbf{Language-grounded signed networks (Section~\ref{sec:structural_polarization}):}
    We convert conversational exchanges into confidence-weighted signed ties, linking observed discourse directly to the agreement and disagreement structure on which polarization is measured. We also retain a human-labelled construction as an LLM-free reference for the resulting structural patterns.

    \item \textbf{Validated structural polarization (Section~\ref{sec:benchmark_validation}):}
    We systematically compare spectral and frustration-based polarization measures on synthetic and real signed networks. While both families can recover planted two-camp structure, their score values differ in their sensitivity to graph size and density. We propose a cycle-space-normalized frustration index, using the cyclomatic number as the normalizing capacity, which provides a more stable basis for cross-window comparison. We introduce both a weighted form that incorporates edge magnitudes and a complementary sign-only count form.

    \item \textbf{Discourse--structure dynamics (Section~\ref{sec:linking_polarization_to_language}):}
    We show that scalar-intensity prevalence is positively associated with count-based structural polarization, with supporting evidence from human-labelled networks, while the weighted and count indices reveal different aspects of the language--structure relationship. We further distinguish contemporaneous association from temporal prediction, finding little evidence of one-month-ahead forecasting but useful information from language and structure measured within the same window.
\end{itemize}

\section{Related work}

We situate our approach by reviewing prior work on network polarization measures, discourse-based NLP signals, and methods that connect text to graph structure.

\subsection{Polarization in social networks}
\label{sec:polarization in social networks}

Polarization describes the emergence of antagonistic camps with dense within-group ties and weak or negative cross-group ties, often reflected in structural separation and bimodal opinion patterns rather than only increased extremism \citep{santos2021link, racz2023towards}. Online polarization is closely tied to echo chambers, where selective exposure and homophilic interaction can reinforce existing views \citep{cinelli2021echo}. Cross-platform studies further show that the strength and form of these patterns vary across platforms and communities, with strong clustering on Facebook and Twitter and more cross-cutting interaction in some Reddit settings \citep{cinelli2021echo, avalle2024persistent}.

To quantify polarization, prior work often constructs interaction graphs, partitions users into groups, and measures separation using random-walk, cut-based, or related structural statistics \citep{garimella2018quantifying}. Such approaches allow polarization to be studied across topics and over time. Signed-network methods make agreement and disagreement explicit rather than treating every interaction as equivalent. Spectral approaches such as Eigen-Sign and SPONGE identify signed communities by exploiting the pattern of positive and negative ties \citep{bonchi2019discovering, cucuringu2019sponge}. A complementary family of methods is based on structural balance and frustration, where the frustration index measures the minimum inconsistency remaining under a two-camp partition \citep{doreian2009partitioning, aref2018measuring}.

These approaches differ not only in how they recover group structure but also in how their scores respond to graph size, density, sparsity, and edge weights. This matters in temporal settings, where the number of active users and interactions can change substantially from one window to another. Prior work has focused mainly on detecting or characterising polarised structure, while the comparability of signed-network scores across changing graph substrates has received less attention. Our work therefore evaluates spectral and frustration-based measures before using them to study temporal relationships between discourse and structural polarization.

\subsection{NLP signals in social media discourse}
\label{sec:NLP in social media}

NLP is widely used to characterise opinions and behaviour in social media, including work that combines text analysis with network perspectives \citep{bail2016combining, guo2024medfluencer}. Classic subjectivity and sentiment methods distinguish evaluative or affective language from factual content and estimate its polarity \citep{montoyo2012subjectivity, cambria2017practical}. Lexicon-based tools such as LIWC provide interpretable measures of affect and psychological language \citep{boyd2022development}, while large-scale studies have examined how emotional expression spreads through online interaction \citep{kramer2014experimental}. The variability of online language has also motivated learned models that move beyond fixed lexical representations \citep{pang2002thumbs}.

Beyond sentiment, toxicity detection targets hostile or disrespectful language, including more implicit forms of toxicity \citep{gevers2022linguistic, wen2023unveiling}, while also raising concerns about bias in automated moderation systems \citep{lee2024people}. Predictability-based measures such as perplexity provide a complementary view of linguistic regularity, while related entropy-based measures have been used to study changes in discourse diversity and alignment \citep{rosen2025antisemitic}. Scalar intensity and hyperbole capture another dimension of discourse: the degree to which speakers strengthen, exaggerate, or intensify propositions \citep{kao2014nonliteral, troiano2018computational}. Such linguistic phenomena are also relevant to misinformation and claim-verification settings \citep{thorne2018fever}.

In this paper, we treat scalar intensity, toxicity, and perplexity as interpretable window-level properties of discourse and examine how they relate to structural polarization measured on signed interaction networks. In particular, we distinguish the broad prevalence of scalar-intensity language from only its extreme tail, allowing us to test where along the intensity distribution the language--structure relationship appears.

\subsection{Connecting text and graph structure}

Recent surveys describe several ways of combining language models and graph learning, including using Large Language Models (LLMs) to support Graph Neural Networks (GNNs), using graphs to support language-model reasoning, and aligning graph and text representations \citep{li2024survey}. Examples include prompting or fine-tuning language encoders to produce graph-relevant representations \citep{he2023harnessing, chien2021node}, reasoning over dynamic or serialised graph structures with LLMs \citep{guo2023gpt4graph, zhang2024llm4dyg}, and methods for aligning textual and graph representations \citep{edwards2021text2mol}. Social-media applications also combine text and interaction structure to study disagreement and community dynamics, including work on climate discourse \citep{su2024decoding}. STEntConv constructs a signed user--entity graph from text-derived stances toward named entities and uses a signed graph convolutional network to predict disagreement between Reddit comment--reply pairs \citep{lorge-etal-2024-stentconv}.

Our focus differs from graph--text integration aimed primarily at improving a single downstream predictor. Instead, language enters our framework in two complementary roles. At the interaction level, a stance model converts conversational exchanges into signed agreement and disagreement ties, providing the network on which structural polarization is measured. At the window level, the same discourse is summarised through interpretable language features whose temporal variation can then be related to the structural polarization series. This positions our work as an empirical bridge between observed discourse and signed-network structure, while keeping the linguistic and structural components separately interpretable.

\section{Problem formulation}
\label{sec:problem_setting}

\paragraph{Temporal setting and notation.}
We consider a sequence of time windows indexed by $t \in \{1,\dots,T\}$, using monthly windows in this study. Each window $t$ yields a signed interaction graph $G_t$ and a scalar structural polarization score $P_t$, computed on the largest connected component of $G_t$. We relate $P_t$ to language signals computed from the same window and examine whether these signals are associated with structural polarization contemporaneously or provide information about polarization at a one-month lead. The framework is evaluated in two settings: \textbf{(i) synthetic benchmarks}, where signed networks are generated under controlled conditions with planted two-camp structure, and \textbf{(ii) real conversational data}, where signed networks are constructed from discussion logs using a windowed pipeline.

Appendix~\ref{sec:appendix_illustrative_pairs} provides fabricated examples of agreement, disagreement, and neutral comment--reply pairs of the kind scored by the stance model.

\paragraph{Three tasks.}
The rest of the paper addresses three tasks on the windowed data. 
\textbf{Task~1 (structural polarization over time):} identify a structural polarization measure $P_t$ that is sensitive to signed two-camp structure while remaining suitable for comparison across windows of different size and density. We evaluate spectral and frustration-based alternatives in Sections~\ref{sec:structural_polarization} and~\ref{sec:benchmark_validation} before selecting the measures used in the language analysis.

\textbf{Task~2 (association):} test whether $P_t$ is associated with a window-level language vector $X_t \in \mathbb{R}^d$ computed from the same conversational content, including scalar-intensity, toxicity, and perplexity features, and examine whether the association remains after adjustment for other language dimensions and graph-level covariates.

\textbf{Task~3 (forecasting and nowcasting):} test whether language provides information about $P_t$ beyond a structure-only baseline based on $P_{t-1}$. We distinguish one-month-ahead forecasting, using $X_{t-1}$, from contemporaneous nowcasting, using $X_t$.
\section{Method}
\label{sec:method}

\begin{figure*}[!t]
  \centering
  \includegraphics[width=\linewidth]{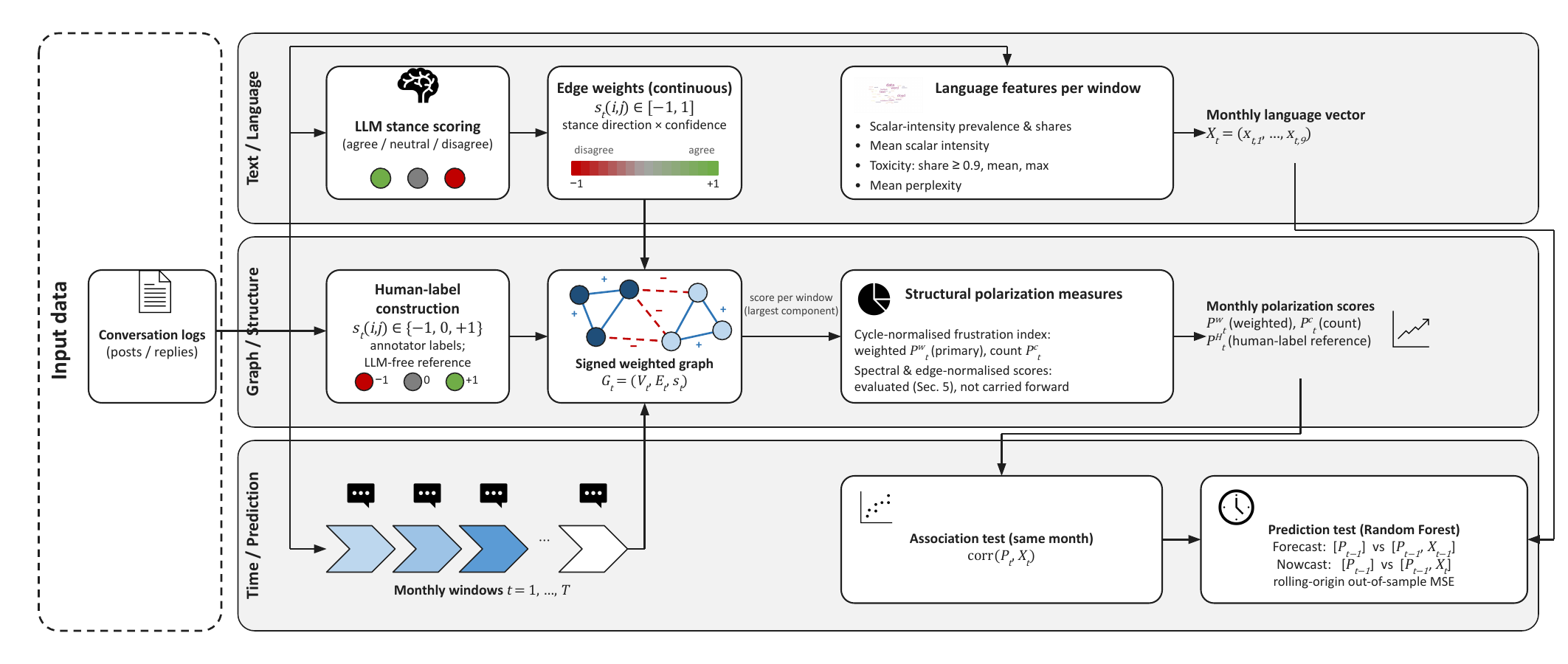}
  \caption{Overview of the analysis pipeline. Language enters in two roles: interaction-level stance scoring defines the signed network on which structural polarization is measured, while window-level language features form the vector $X_t$. Candidate spectral and frustration-based measures are evaluated before the cycle-normalized frustration indices are used in the language analysis. A human-labelled construction provides an LLM-free reference. Downstream analyses examine contemporaneous association, covariate adjustment, one-month-ahead forecasting, and contemporaneous nowcasting.}
  \label{fig:workflow}
\end{figure*}

\subsection{Workflow overview}
\label{sec:workflow_overview}

Figure~\ref{fig:workflow} summarises the pipeline. Conversation logs are segmented into monthly windows and converted into signed interaction networks. We first evaluate candidate structural polarization measures for cross-window comparison (Section~\ref{sec:benchmark_validation}), then pair the retained polarization series with window-level language features for association, adjustment, and prediction analyses.

\subsection{Structural polarization}
\label{sec:structural_polarization}

\subsubsection{Windowed signed interaction graphs}

\paragraph{Corpus.}
We analyse the r/Brexit portion of DEBAGREEMENT \citep{pougue2021debagreement}, a human-annotated comment--reply dataset. We retain the 5{,}555 of 15{,}745 r/Brexit pairs on which all annotators agreed. This agreement criterion selects the clearer exchanges for both graph constructions below. Interactions are grouped into 30 monthly windows from December 2018 to May 2021.

For each window $t$, we construct an undirected signed graph
$
G_t=(V_t,E_t,s_t),
$
where $V_t$ denotes the set of active users, $E_t$ the set of interacting user pairs, and $s_t$ assigns an aggregated signed relation in $[-1,1]$ to each interacting pair. We consider two graph constructions.

In the \textbf{continuous construction}, each interaction receives a stance-model weight $w\in[-1,1]$, reflecting both stance direction and confidence, and the pair weight is the mean of $w$ across interactions between that pair. In the \textbf{human-label construction}, interactions are assigned $-1/0/+1$ from the corpus's human disagreement, neutral, and agreement labels, and the pair weight is the sign of their mean. The human-label construction provides an LLM-free reference. Structural scores are computed on the largest connected component of each graph.

\paragraph{Stance scoring.}
\label{sec:discrete_continuous_weights}

We LoRA-adapt \texttt{meta-llama/Llama-3.2-1B} to a three-class stance task over
$\{\texttt{disagree},\texttt{neutral},\texttt{agree}\}$. Given the predicted class probabilities, each interaction receives the signed weight
$
w=
(p_{\text{agree}}-p_{\text{disagree}})
(1-p_{\text{neutral}}).
$
The first term determines stance direction, while the second reduces the magnitude of neutral or uncertain interactions. The stance classifier achieves a weighted F1 score of 0.85 on the held-out DEBAGREEMENT test set. Full training, validation, and data-overlap details are reported in Appendix~\ref{sec:appendix_stance_scoring}. Because the stance model is trained on data that overlap the analyzed corpus, the continuous construction is not independent of the training labels; the human-label construction therefore provides an LLM-free reference. These signed relations summarize local agreement and disagreement in observed comment–reply exchanges. Disagreement with a particular statement does not necessarily imply opposing political identities. Edge magnitude reflects the stance model’s output probabilities and should not be interpreted as social tie strength or influence.

\subsubsection{Polarization measures}

We consider two families of signed-network polarization measures: spectral measures, which quantify global signed alignment through an eigenvalue objective, and frustration-based measures, which quantify how closely the graph can be partitioned into two internally consistent camps. We evaluate both families on synthetic and real-network benchmarks in Section~\ref{sec:benchmark_validation} before selecting the measures used in the language analysis. 

Throughout this paper, structural polarization refers to the consistency of observed agreement and disagreement relations with a two-camp signed partition. We use this formulation as a common structural criterion across windows, without assuming that the community consists of exactly two stable political groups. The inferred camps are analytical partitions; we do not interpret them as established Leave/Remain identities or as direct measures of users’ ideological positions.

\paragraph{Spectral measures.}
Following Eigen-Sign \citep{bonchi2019discovering}, let $A^{(t)}$ denote the signed weighted adjacency matrix in window $t$ and let $x$ be the leading-eigenvector assignment used to recover the two-camp partition. The normalized Eigen-Sign score is $P^{\mathrm{eig}}_t=(x^\top A^{(t)}x)/((x^\top x)W_t)$, where $W_t=\sum_{(i,j)\in E_t}|A^{(t)}_{ij}|$ is the total absolute edge weight.

We also consider a degree-normalized formulation based on the generalised Rayleigh quotient. Let $\bar D_{ii}=\sum_j|A^{(t)}_{ij}|$ denote the absolute weighted degree of node $i$. We define $P^{\mathrm{gen}}_t=\lambda_{\max}(\bar D^{-1/2}A^{(t)}\bar D^{-1/2})$, which is bounded in $[-1,1]$, with $1$ corresponding to a perfectly balanced signed graph. Section~\ref{sec:benchmark_validation} evaluates the suitability of both spectral scores for cross-window comparison.

\paragraph{Frustration-based measures.}
Frustration measures the amount of inconsistency remaining after the nodes have been optimally assigned to two camps, maximizing with-camp agreement and across-camp disagreement. For a graph with $n_t$ nodes, $m_t$ non-neutral edges, and $c_t$ connected components, the cyclomatic number is $\kappa_t=m_t-n_t+c_t$, the dimension of the graph's cycle space. Because all scores are computed on the largest connected component, $c_t=1$ and $\kappa_t=m_t-n_t+1$.

Let $F_t$ be the minimum number of frustrated edges over binary assignments $x\in\{\pm1\}^{n_t}$, where an edge $(i,j)$ is frustrated when $A^{(t)}_{ij}x_ix_j<0$. The standard edge-normalized frustration index \citep{aref2018measuring}, $P^{\mathrm{edge}}_t=1-2F_t/m_t$, normalizes by the bound $F_t\le m_t/2$. Since frustration arises only through incompatible cycle constraints and a forest is always balanced, we instead normalize by cycle-space capacity and define the count index as $P^{\mathrm{c}}_t=1-F_t/\kappa_t$.

To retain the magnitude of the continuous stance weights, we also define a weighted form. Let $F^{\mathrm{w}}_t$ be the minimum total absolute weight carried by frustrated edges, and let $C^{\mathrm{w}}_t=W_t-\mathrm{maxforest}_t$, where $\mathrm{maxforest}_t$ is the total weight of a maximum-weight spanning forest. We define $P^{\mathrm{w}}_t=1-F^{\mathrm{w}}_t/C^{\mathrm{w}}_t$. When all non-neutral edges have unit magnitude, $P^{\mathrm{w}}_t$ reduces exactly to $P^{\mathrm{c}}_t$. Thus, the count index captures sign direction only, while the weighted index additionally retains stance-model confidence.

Section~\ref{sec:benchmark_validation} compares all five candidate scores. We carry $P^{\mathrm{w}}_t$ and $P^{\mathrm{c}}_t$ into the language analysis, and denote the count index on the human-label construction by $P^{\mathrm{H}}_t$.
\subsubsection{Analysis sample}
\label{sec:sample_rule}

A monthly window is retained when the frustration index is defined on the continuous graph and the human-label graph contains at least three independent cycles, avoiding control scores determined by only one or two cycle constraints. This yields 24 of the 30 monthly windows, spanning January 2019 to January 2021 with one interior gap in November 2019. The same 24 windows are used throughout the real-data analyses. The excluded windows, the sign-shuffle rationale for the sample rule, and sensitivity to low-cycle windows are reported in Appendix~\ref{sec:appendix_sample_rule}.

\subsection{Language features}
\label{sec:language_features}

We extract interpretable language signals capturing scalar intensity, toxicity, and predictability. All features are computed on deduplicated parent--child comment pairs within each monthly window and aggregated into a window-level vector $X_t$ aligned with the structural polarization score $P_t$.

\paragraph{Scalar intensity.}
We generate a scalar-intensity lexicon five times using the same calibrated GPT-5 prompt. For each interaction, we record the maximum intensity of any matched lexicon term, assigning $0$ when no term is present. We then compute the shares of interactions reaching thresholds $\theta\in\{0^{+},0.5,0.7,0.9\}$ and the mean maximum intensity. The share at $\theta=0^{+}$, $x^{\mathrm{any}}_t$, measures scalar-intensity prevalence, while $x^{\mathrm{scalar}}_t$ denotes the extreme share at $\theta=0.9$. Feature values are averaged across the five lexicon generations. Full lexicon construction and robustness analyses are reported in Appendix~\ref{sec:scalar_lexicon}.

\paragraph{Toxicity.}
Each interaction is scored with \texttt{unitary/toxic-bert}. We compute the share of high-toxicity interactions, together with the monthly mean and maximum toxicity score.

\paragraph{Perplexity.}
We measure discourse predictability using the monthly mean of per-interaction perplexity under the pretrained GPT-2 base model. Higher values indicate less predictable language.
\subsection{Association and prediction design}
\label{sec:prediction_design}

We examine three relationships between language and structural polarization: contemporaneous association, covariate-adjusted association, and prediction at lag~1 and lag~0. Throughout, $P_t$ denotes either $P^{\mathrm{w}}_t$ or $P^{\mathrm{c}}_t$ on the continuous graph, or $P^{\mathrm{H}}_t$ on the human-label graph.

\paragraph{Contemporaneous association.}
For each language feature, we report Pearson $r$ and Spearman $\rho$ with $P_t$ across the 24 analysed months. To account for temporal dependence, we complement nominal correlation tests with a circular time-shift test that preserves each series' autocorrelation while breaking their alignment. For Pearson correlation, the feature series is rotated through all 23 non-zero offsets and
$
p_{\mathrm{shift}}
=
\frac{
1+
\sum_{k=1}^{23}
\mathbb{I}
\left(
|r_k|\ge|r_{\mathrm{obs}}|
\right)
}{24}.
$
The minimum attainable value is $1/24\approx0.042$. Holm \citep{holm1979simple} and Benjamini--Hochberg \citep{benjamini1995controlling} corrections over the nine-feature family are applied to the nominal $p$-values for each structural measure.

\paragraph{Multivariable adjustment.}
We use ordinary least-squares models to test whether the association between scalar prevalence and $P_t$ remains after accounting for other factors. We adjust for language covariates (mean toxicity and mean perplexity), graph covariates (node count and mean degree), and both sets jointly. We report the partial association of scalar prevalence with $P_t$ together with a conditional circular-shift test. Full model specifications and results are provided in Appendix~\ref{sec:appendix_adjustment}.

\paragraph{Forecasting and nowcasting.}
We compare a structure-only Random Forest based on $P_{t-1}$ with an augmented model that additionally includes scalar prevalence. In the one-month-ahead forecasting analysis, the augmented model uses $x^{\mathrm{any}}_{t-1}$; in the contemporaneous nowcasting analysis, it uses $x^{\mathrm{any}}_t$.

The 24 retained months yield 22 consecutive-month pairs because of the November 2019 gap. Our primary evaluation uses a rolling-origin design: after an initial 10-pair burn-in, each subsequent month is predicted using models fitted only to earlier observations, yielding 12 out-of-sample evaluation months. Performance is measured using mean squared error (MSE), with relative improvement defined as
$
\mathrm{Gain}
=
1-
\frac{
\mathrm{MSE}_{\mathrm{aug}}
}{
\mathrm{MSE}_{\mathrm{base}}
}.
$
Prediction gains are evaluated against circular shifts of the language feature sequence. Random Forest settings, the chronological 80/20 comparison, training-fold feature selection, and month-level diagnostics are reported in Appendix~\ref{sec:appendix_adjustment}.

\section{Benchmarking and Validation of Polarization Measures}
\label{sec:benchmark_validation}

Before relating structural polarization to language, we evaluate whether the candidate measures capture the intended signed two-camp structure and whether their scores can be compared across changing interaction networks. We consider the two spectral measures, normalized Eigen-Sign $P^{\mathrm{eig}}_t$ and generalised spectral $P^{\mathrm{gen}}_t$, together with edge-normalized frustration $P^{\mathrm{edge}}_t$ and the cycle-normalized count and weighted frustration indices, $P^{\mathrm{c}}_t$ and $P^{\mathrm{w}}_t$.

\subsection{Synthetic benchmarks}
\label{sec:synthetic_benchmarks}

\paragraph{Benchmark design.}
We first evaluate the measures on two synthetic graph families with planted two-camp structure: signed stochastic block models (SBM) and signed Barabási--Albert (BA) networks. We vary the amount of sign noise and graph density, while keeping the planted groups hidden from all methods. The SBM provides a controlled community structure, while the BA model introduces degree heterogeneity and, in its continuous form, varying edge magnitudes. Full generator specifications and parameter settings are reported in Appendix~\ref{sec:appendix_synthetic_generators}.

\paragraph{Sensitivity to planted structure.}
All five measures show an overall negative Spearman correlation with sign noise $\eta$, with the cycle-normalized count index $P^{\mathrm{c}}$ and weighted index $P^{\mathrm{w}}$ showing the strongest relationship across both graph families. On the SBM benchmark, both achieve $\rho=-0.77$, compared with $-0.47$ to $-0.60$ for the alternatives (Table~\ref{tab:validity}, Sensitivity column). Full results, including the BA benchmarks and variation with graph density, are reported in Appendix~\ref{sec:appendix_synthetic_generators}.

\paragraph{Recovery of planted structure.}
Across both synthetic families, the spectral and frustration-based approaches behave consistently when a recoverable two-camp signal is present. At zero sign noise they recover the same balanced partition, and their agreement decreases as sign noise weakens the planted structure. Importantly, denser graphs remain consistent for longer: with more structural information available, both methods recover the planted camps more reliably.

For example, on the SBM at low sign noise, the generalised spectral and weighted-frustration partitions achieve similar agreement with the planted labels: ARI is approximately $0.61/0.64$ at mean degree $3$, $0.95/0.93$ at degree $6$, and $0.99/1.00$ at degree $12$. The same pattern appears on the BA networks, where agreement remains high at low noise and persists longer as graph density increases.

\begin{figure*}[htbp]
    \centering
    \includegraphics[width=0.8\linewidth]{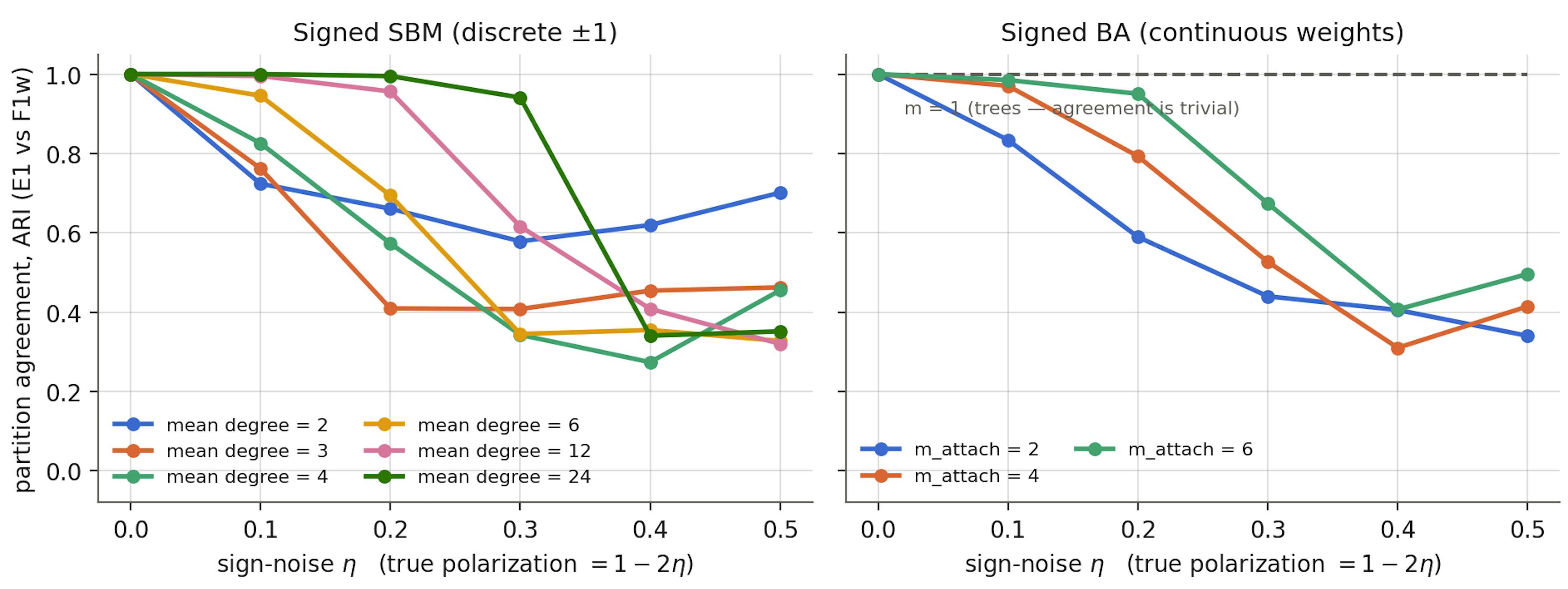}
    \caption{Partition agreement (ARI) between the generalised-spectral and weighted-frustration solutions as sign noise $\eta$ increases. The left panel shows signed SBM graphs and the right panel signed BA graphs with continuous weights. Agreement is strongest when the planted signal is clear, and denser graphs remain aligned for longer as noise increases. The $m_{\mathrm{attach}}=1$ BA setting is a degenerate tree case in which frustration cannot arise.}
    \label{fig:alignment_synthetic}
\end{figure*}

Figure~\ref{fig:alignment_synthetic} shows the agreement between the generalised-spectral and weighted-frustration partitions across both generators. The two approaches are therefore broadly consistent as methods for identifying two-camp structure; their differences become more important when we consider how the resulting \emph{scores} behave across networks.

% The score values show the same broad response to increasing sign noise: all candidate measures decrease as the planted structure is degraded. The cycle-normalized frustration measures show the strongest and most consistent sensitivity across the two graph families. On paired BA graphs, the weighted index on continuous edges and the count index on sign-only copies are also closely aligned, indicating that incorporating edge magnitudes preserves the underlying structural signal. Detailed sensitivity, recovery, and continuous-versus-discrete results are reported in Appendix~\ref{sec:appendix_synthetic_generators}.

\subsection{Real-network validation}
\label{sec:real_sanity}

The synthetic experiments show that both spectral and frustration-based approaches can recover meaningful signed structure. For the language analysis, however, we additionally require the \emph{score itself} to be comparable across monthly networks whose numbers of users and interactions change over time. We therefore test each score against node count, edge count, mean degree, and cyclomatic number on the 24 analyzed Reddit networks.

\paragraph{Dependence on graph structure.}
The real networks reveal important differences between the measures. The normalized Eigen-Sign score is strongly associated with graph size, including a Spearman correlation of
$
\rho=-0.989
$
with edge count. Its monthly variation therefore reflects interaction volume as well as signed structure.

\begin{figure*}[h]
    \centering
    \includegraphics[width=\linewidth]{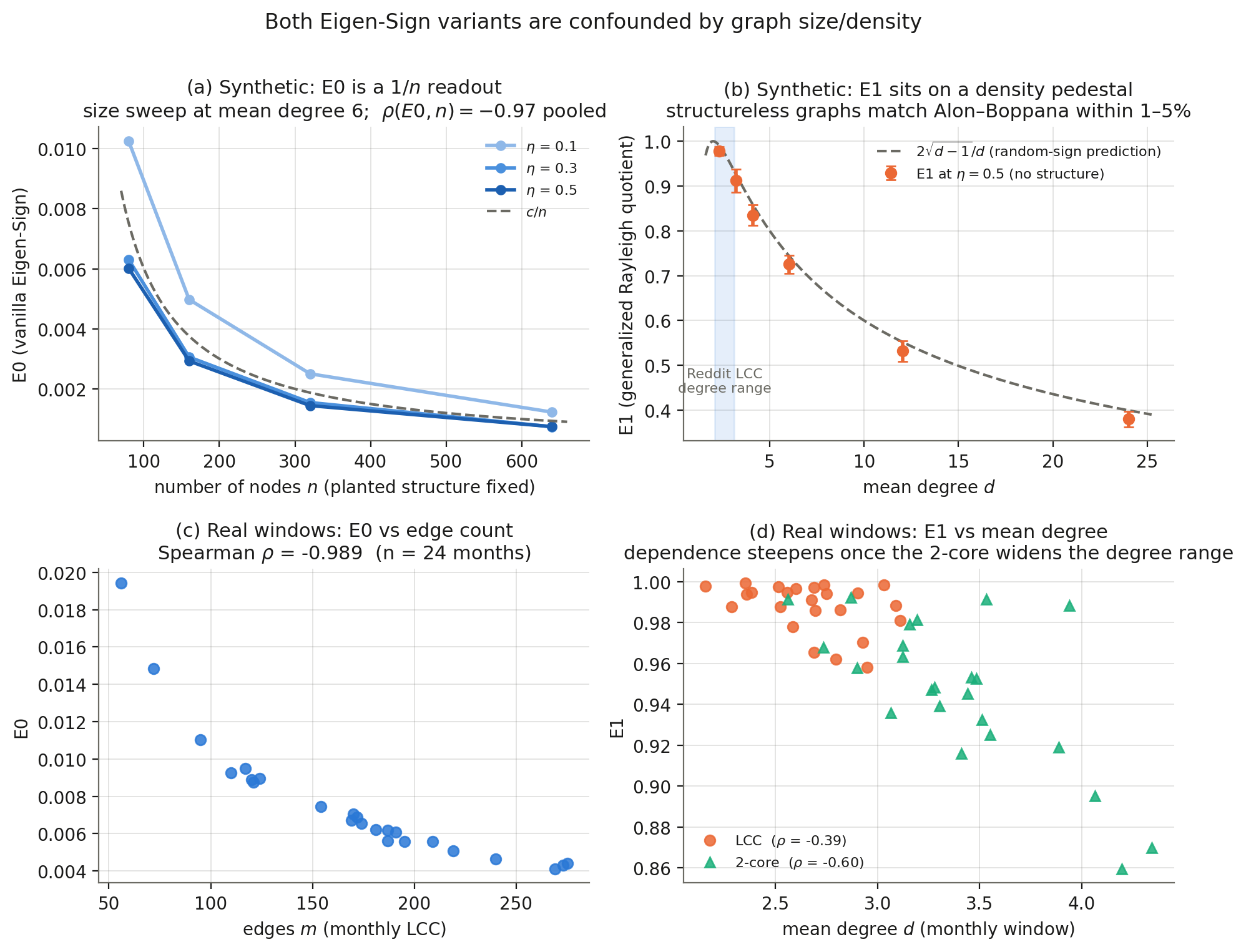}
    \caption{Size and density effects in the spectral measures. Controlled analyses show the size dependence of normalized Eigen-Sign and the density-dependent baseline of the generalised spectral score; the real monthly networks exhibit the corresponding relationships with graph size and mean degree. Here E0 denotes normalized Eigen-Sign \(P^{\mathrm{eig}}\), and E1 denotes the generalised spectral score \(P^{\mathrm{gen}}\).}
    \label{fig:eigen_confounds}
\end{figure*}

The generalised spectral formulation substantially reduces this direct size dependence, but remains sensitive to graph density and substrate. It passes the size/density screen on the largest connected component but fails when the same analysis is repeated on the 2-core. Figure~\ref{fig:eigen_confounds} illustrates the size dependence of normalized Eigen-Sign and the density-dependent baseline of the generalised spectral score.

A similar issue appears for edge-normalized frustration, which is strongly associated with mean degree,
$
\rho=-0.941.
$
By contrast, the weighted cycle-normalized frustration index has the smallest maximum absolute correlation with the tested graph characteristics ($0.246$, $p_{\mathrm{fw}}=0.394$; Table~\ref{tab:screen} in Appendix~\ref{sec:appendix_real_sanity}). Its month ranking is unchanged between the largest connected component and the 2-core ($\rho=1.000$; see ``Substrate robustness'' in the same appendix). The count index behaves similarly, although it retains a moderate dependence on mean degree and lies close to the boundary of the size/density screen.

\paragraph{Agreement and reliability.}
The real-network results reinforce the distinction between recovering a partition and obtaining a comparable cross-window score. The spectral and frustration-based methods can identify similar two-camp structure in controlled settings, but their monthly score series on the real networks are not interchangeable. The weighted and count frustration indices also differ in how they rank some months, reflecting the additional information carried by edge magnitudes.

Among the frustration measures, the weighted index is more stable across graph substrates and aligns more closely with the LLM-free human-label reference. The count index is retained as a complementary sign-only view. Full size/density screens, reliability analyses, and optimization checks are reported in Appendix~\ref{sec:appendix_real_sanity}.

\subsection{Measure selection}

Table~\ref{tab:validity} summarises the validation results. \textit{Screen} assesses associations with graph size and density using
$
M(P)=\max_{z\in\{n,m,\bar{d},\kappa\}}|\rho_{\mathrm{S}}(P,z)|,
$
where $\rho_{\mathrm{S}}$ is Spearman correlation, and $n$, $m$, $\bar{d}$, and $\kappa$ denote node count, edge count, mean degree, and cyclomatic number. A family-wise $p$-value is obtained from 5{,}000 shared permutations. The summary considers results on both the largest connected component and the 2-core; full results appear in Appendix~\ref{sec:appendix_real_sanity}, under ``Size and density screen'' and ``Substrate robustness''.

\textit{Spread} measures score variation across SBM density settings at $\eta=0.5$, where the planted sign signal is absent. \textit{Sensitivity} is the Spearman correlation between the score and $\eta$ across the complete SBM grid. These criteria are detailed in Appendix~\ref{sec:appendix_synthetic_generators}, under ``Specificity to density and graph size'' and ``Sensitivity to planted sign structure''. \textit{ICC} is the intraclass correlation under 40 repeated draws retaining 95\% of edges in each monthly network; see Table~\ref{tab:reliability} in Appendix~\ref{sec:appendix_real_sanity}.

\begin{table}[t]
\centering
\small
\setlength{\tabcolsep}{1pt}
\begin{tabular}{lcccc}
\toprule
& \textbf{Real} & \multicolumn{2}{c}{\textbf{Synthetic SBM}} & \textbf{Real} \\
\cmidrule(lr){2-2}\cmidrule(lr){3-4}\cmidrule(lr){5-5}
Score & Screen & Spread & Sensitivity & ICC \\
\midrule
Norm.\ Eigen-Sign $P^{\mathrm{eig}}$
& fail & $0.011^{*}$ & $-0.47$ & $0.99$ \\

Gen.\ spectral $P^{\mathrm{gen}}$
& fail & $0.598$ & $-0.51$ & $0.79$ \\

Edge-norm.\ frustr.\ $P^{\mathrm{edge}}$
& fail & $0.607$ & $-0.60$ & $0.95$ \\

Count frustr.\ $P^{\mathrm{c}}$
& borderline & $0.087$ & $-0.77$ & $0.71$ \\

Weighted frustr.\ $P^{\mathrm{w}}$
& pass & $0.087$ & $-0.77$ & $0.70$ \\
\bottomrule
\end{tabular}
\caption{Validation summary for the five structural polarization measures. Screen and ICC are computed on the real monthly networks, while spread and sensitivity are computed from the signed-SBM grid. $^{*}$The small absolute spread of normalized Eigen-Sign reflects its compressed size-dependent scale.}
\label{tab:validity}
\end{table}

The cycle-normalized frustration measures provide the strongest overall basis for cross-window comparison. Both show low density-driven variation and strong sensitivity to planted structure in the synthetic experiments. On the real networks, the weighted index shows little dependence on graph size or density, while the count index is borderline because of its association with mean degree. We therefore use the weighted index as the primary structural measure and retain the count index as a complementary sign-only view.

The spectral measures are not carried into the language analysis. Although they recover planted two-camp structure in the synthetic experiments, their score values are less stable across changing graph substrates: normalized Eigen-Sign is strongly related to graph size, while the generalised spectral score remains sensitive to density. Edge-normalized frustration shows a similar density dependence.

\paragraph{Absolute level relative to random signs.}
The cycle-normalized frustration scores have a non-zero, month-specific random-sign baseline. We therefore interpret them primarily through relative variation across months rather than as absolute evidence that each monthly network is polarised beyond chance. Full sign-shuffle results are reported in Appendix~\ref{sec:appendix_sample_rule}.
\section{Linking Structural Polarization to Language}
\label{sec:linking_polarization_to_language}

Having established the structural polarization measures, we next examine how they relate to window-level discourse. All analyses use the same 24 monthly windows defined in Section~\ref{sec:sample_rule}. We report the weighted frustration index $P^{\mathrm{w}}_t$ as the primary validated structural measure, the count index $P^{\mathrm{c}}_t$ as a complementary sign-only view, and the human-label score $P^{\mathrm{H}}_t$ as an LLM-free reference.

\subsection{Association between structural polarization and language}
\label{sec:assoc_results}

The clearest language relationship is with the prevalence of scalar-intensity vocabulary. The share of interactions containing at least one scalar term is positively associated with the count frustration index,
$
r=+0.56,
$
and exceeds every circular time shift (unadjusted $p_{\mathrm{shift}}=0.042$). The same relationship is weaker for the weighted index ($r=+0.36$), while the human-label reference shows a similar positive pattern in rank. Full association results, including Spearman correlations and multiplicity corrections, are reported in Appendix~\ref{sec:appendix_adjustment}. Threshold sensitivity and lexicon robustness are reported in Appendix~\ref{sec:scalar_lexicon}.

\begin{figure*}[t]
  \centering
  \includegraphics[width=\linewidth]{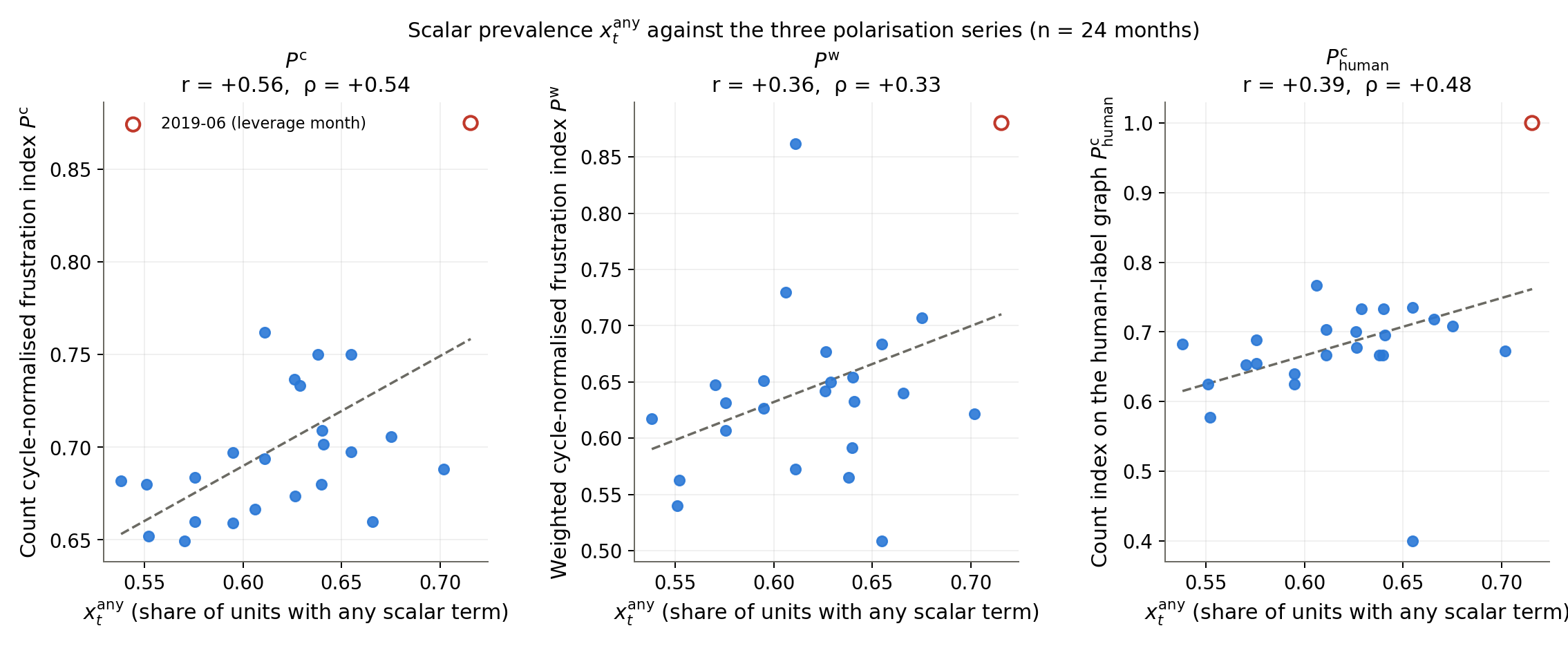}
  \caption{Scalar-intensity prevalence $x^{\mathrm{any}}_t$ against the count index (left), weighted index (centre), and human-label reference (right), one point per analyzed month. The ringed point is June 2019, which is examined separately as a leverage check. The human-label quantity shown as \(P^{c}_{\mathrm{human}}\) in the figure corresponds to \(P^{H}\) in the main text.}
  \label{fig:assoc_language_vs_polarization}
\end{figure*}

\paragraph{Other language dimensions.}
Toxicity and perplexity show no comparable association with either the count or weighted frustration index. These features are therefore retained mainly as adjustment variables rather than as primary findings.

\paragraph{Multi-factor adjustment and sensitivity.}
We next test whether the association between scalar prevalence and the count index remains after accounting for other factors. In the language-adjusted model, we control for mean toxicity and mean perplexity; in the graph-adjusted model, we control for node count and mean degree; and in the combined model, we include both sets of covariates. The scalar-prevalence association remains similar after adjustment for the language variables, but becomes less certain after accounting for graph size and density. The result is also sensitive to June 2019: removing this month weakens the Pearson association, although the positive rank relationship remains. We therefore interpret the finding as a descriptive cross-month association. Full multi-factor and leverage analyses are reported in Appendix~\ref{sec:appendix_adjustment}.

\subsection{Lagged and contemporaneous prediction}
\label{sec:rf_prediction_results}

We next ask whether scalar prevalence contributes information beyond the previous month's structural score. We distinguish one-month-ahead forecasting, using the previous month's language, from contemporaneous nowcasting, using language from the month whose structural score is being estimated.

\paragraph{Forecasting and nowcasting.}
We find no evidence that the previous month's scalar prevalence improves prediction of the following month's polarization. This result is consistent across the count, weighted, and human-label series under rolling-origin evaluation and circular-shift testing, suggesting little month-to-month predictive signal at this resolution. In contrast, current-month scalar prevalence provides useful contemporaneous information: for the count index, adding it to a structure-only model reduces rolling-origin out-of-sample MSE by approximately $74\%$, and the improvement is recovered when the language feature is selected within each training fold. The corresponding improvement for the weighted index is smaller and less stable.
\begin{table}[t]
\centering
\small
\begin{tabular}{lccc}
\toprule
Target & Lag & Gain & $p_{\mathrm{shift}}$ \\
\midrule
Count $P^{\mathrm{c}}$ & $t-1$ & $+8.4\%$ & $.95$ \\
Weighted $P^{\mathrm{w}}$ & $t-1$ & $+6.6\%$ & $.36$ \\
Human $P^{\mathrm{H}}$ & $t-1$ & $-2.1\%$ & $.41$ \\
\midrule
Count $P^{\mathrm{c}}$ & $t$ & $\mathbf{+74.2\%}$ & $\mathbf{.045}$ \\
Weighted $P^{\mathrm{w}}$ & $t$ & $+23.8\%$ & $.045$ \\
\bottomrule
\end{tabular}
\caption{Rolling-origin prediction results. Lag $t-1$ tests one-month-ahead forecasting, while lag $t$ uses contemporaneous scalar prevalence. Detailed MSE values, within-fold feature selection, and chronological-split results are reported in Appendix~\ref{sec:appendix_adjustment}.}
\label{tab:lagged_rf_results}
\end{table}

The language--structure relationship is therefore primarily contemporaneous at the monthly scale examined here. Scalar prevalence contains useful information about the count-based structural score within the same month, but we find no evidence that it precedes changes in polarization by one month. We make no causal or temporal-precedence claim from the contemporaneous result.

\section{Discussion and Conclusion}
\label{sec:discussion}

\subsection{Two roles of language}
\label{sec:discussion_continuous}

Our framework uses language in two complementary ways. At the interaction level, conversational exchanges are converted into signed relations that define the network on which structural polarization is measured. At the window level, the same discourse is summarised through interpretable language features that can be related to changes in structural polarization over time. This provides a direct bridge between observed text and signed interaction structure while keeping the two levels of analysis separately interpretable.

Because the signed networks and window-level language features are derived from the same conversational text, their associations may partly reflect shared linguistic cues. The human-labelled construction provides a comparison without the language model stance scorer, but both constructions use the same source material. It therefore does not rule out shared-text dependence. We interpret the observed associations and nowcasting gains as descriptive relationships within this measurement framework, rather than evidence that language causes changes in independently measured political polarization.

\paragraph{Choosing a structural polarization measure.}
The validation results show that recovering planted two-camp structure in synthetic benchmarks is not sufficient to establish a score’s suitability for temporal comparison: the score must also remain comparable as network size and density change.
The cycle-normalized frustration measures are better suited to this setting than the spectral and edge-normalized alternatives. The weighted index provides the more stable structural measure and retains stance-confidence information, while the count index offers a complementary view based only on agreement and disagreement directions.

The distinction between the two frustration measures is also informative. The weighted index aligns more closely with the human-label reference and is better behaved across graph substrates, whereas the count index shows the clearer relationship with scalar-intensity language. This suggests that edge direction and edge magnitude capture related but distinct aspects of signed interaction structure rather than one measure simply replacing the other.

\subsection{Conclusion}

We introduced a language-grounded framework for linking discourse to structural polarization in signed interaction networks. Language is used both to construct agreement and disagreement relations between users and to derive interpretable window-level signals that can be related to the resulting network structure.

Before studying this language--structure relationship, we compared spectral and frustration-based polarization measures on synthetic and real networks. Both families can recover planted two-camp structure, but their scores differ in suitability for cross-window comparison. Cycle-normalized frustration provides the more stable basis for the temporal analysis, with a weighted form that incorporates stance-model confidence and a complementary count form based on edge signs.

On monthly Reddit Brexit discussions, scalar-intensity prevalence is positively associated with the count-based structural score, with a similar rank-level pattern in the human-label reference. The weighted index provides the more stable structural estimate but shows a weaker relationship with this language feature, highlighting the complementary roles of edge direction and edge magnitude.

We find little evidence that language in one month forecasts polarization in the next. Instead, scalar-intensity prevalence provides useful contemporaneous information about structural polarization within the same month. Overall, the results show that discourse and signed interaction structure can be studied within one framework, while also showing that careful validation of the structural measure is essential before interpreting their relationship.
\section*{Limitations}

\paragraph{Binary partitioning.}
Frustration minimization and spectral rounding assume two camps, potentially oversimplifying discussions with multiple factions. Multi-block evaluation, for example with SPONGE, would require larger networks with sufficient interactions within and between groups for adequate statistical power.

\paragraph{Scope and generalizability.}
Our case study is limited to Reddit Brexit and unanimously labelled comment--reply pairs, favouring clearer exchanges. Moderation may shape observable language and interactions, so toxicity or other features could be predictive in less moderated settings. Generalization across platforms, time periods, and political contexts remains untested.

\paragraph{Sample size and exclusions.}
Six of 30 monthly windows are excluded (four forests and two one-cycle human-label graphs), leaving 24 months, 22 consecutive-month pairs, and 12 rolling-origin evaluation months. June 2019 has only three human-graph cycles, all balanced, and is a leverage point: without it, the headline Pearson correlation has $p\approx0.07$ and the rank correlation has $p\approx0.02$. The lag-1 null concerns only monthly resolution; the series cannot support a lag sweep, and finer time scales remain untested.

\paragraph{Sparse regime and optimization.}
Sensitivity to planted structure weakens at the corpus's mean degree ($2$--$3$), while residual density dependence within fixed noise levels cancels only in aggregate. Reported frustration values are numerical upper bounds on the global minima; independent optimization runs differ by one edge for the count measure in the largest windows.

\paragraph{Language features and model judgments.}
We examine scalar-intensity prevalence, toxicity, and perplexity; other discourse features remain untested. Lexicon regeneration assesses sampling robustness under one prompt, not prompt robustness or agreement with human judgments. Studies of subjective surveys report systematic differences between LLM and human responses \citep{santurkar2023whose}, motivating caution about potential biases in our model-derived stance and intensity scores.
\section*{Ethics Statement}

We use the DEBAGREEMENT dataset, which was collected and released by \citet{pougue2021debagreement} under institutional ethics approval. 
This project received ethical approval from the University of Oxford's 
Medical Sciences Interdivisional Research Ethics Committee (MS IDREC) 
under reference R83093/RE002. We analyze publicly available Reddit 
discussions but do not quote user content verbatim, as short excerpts 
can be traced to identifiable users; illustrative examples in the main 
text are fabricated and appendix examples are redacted. The methods 
are intended for descriptive research on group dynamics, not for 
identifying or moderating individual users.

% Since December 2023, a "Limitations" section has been required for all papers submitted to ACL Rolling Review (ARR). This section should be placed at the end of the paper, before the references. The "Limitations" section (along with, optionally, a section for ethical considerations) may be up to one page and will not count toward the final page limit. Note that these files may be used by venues that do not rely on ARR so it is recommended to verify the requirement of a "Limitations" section and other criteria with the venue in question.

\clearpage
\bibliography{custom}

\clearpage
\appendix

% =========================================================
\section{Illustrative stance examples}
\label{sec:appendix_illustrative_pairs}

Table~\ref{tab:illustrative_pairs} shows fabricated comment--reply pairs of the kind scored by our stance model. We do not quote Reddit content verbatim for ethical reasons, as even short excerpts may be traceable to individual users. The examples illustrate agreement, disagreement, and neutral or mixed stance.

\begin{table*}[h]
\centering
\tiny
\renewcommand{\arraystretch}{1.2}
\setlength{\tabcolsep}{6pt}
\begin{tabular}{p{0.30\linewidth} p{0.40\linewidth} p{0.13\linewidth}}
\toprule
\textbf{Comment} & \textbf{Reply} & \textbf{Label} \\
\midrule
It will be worse than anyone expects. The models are totally bogus. Why am I not surprised to see this level of idiocy?
&
That's just the nature of politicians, they'll say anything to stay in power.
&
Agreement \\
\midrule
We don't need a crackdown on immigrants coming here and doing jobs other people don't want. We need to tackle the climate crisis and ecological breakdown!
&
Did you read the article?? It doesn't say we need to reduce immigration, just get a fair deal with other countries.
&
Disagreement \\
\midrule
I think it would be a fair assumption that we won't have so many cheap fresh vegetables if Brexit happens.
&
I see what you mean, imports from Portugal and Spain will probably get more expensive. But on the other hand, imports from Africa might increase once we can control our own trade agreements.
&
Neutral \\
\bottomrule
\end{tabular}
\caption{Illustrative fabricated comment--reply pairs of the kind scored by our stance model, one per class.}
\label{tab:illustrative_pairs}
\end{table*}

% =========================================================
\section{Stance scoring: training and inference details}
\label{sec:appendix_stance_scoring}

This appendix provides the training, inference, and validation details for the LoRA-adapted stance classifier introduced in Section~\ref{sec:discrete_continuous_weights}.

\subsection{Model and adaptation}

We adapt \texttt{meta-llama/Llama-3.2-1B} using low-rank adapters (LoRA) attached to the model's attention and feed-forward projections. Only the adapter parameters are updated, while the base model remains frozen.

\subsection{Training data}

The model is trained on the  DEBAGREEMENT release \citep{pougue2021debagreement}, containing 42{,}894 human-annotated Reddit comment--reply pairs across five subreddits. Each interaction is labelled with one of three stance classes:
\texttt{disagree}, \texttt{neutral}, or \texttt{agree}. The r/Brexit subset analyzed in this paper is drawn from the same release.

Using discourse-level stance annotations rather than weak supervision or sentiment proxies allows the signed interaction weights used in Section~\ref{sec:discrete_continuous_weights} to represent agreement and disagreement rather than affective polarity alone.

\subsection{Training setup}

Stance prediction is formulated as a verbalised three-class task. Given a comment--reply pair, the model predicts a single label token from
$\{\texttt{disagree},\texttt{neutral},\texttt{agree}\}$.
Training minimises cross-entropy over the three label-token logits using a batch size of 4, learning rate $2\times10^{-5}$, five epochs, and a maximum sequence length of 512.

\subsection{Inference}

At inference, the model produces logits
$\ell_{\text{disagree}}$,
$\ell_{\text{neutral}}$, and
$\ell_{\text{agree}}$.
Applying softmax gives the class probabilities

$$
p_i
=
\frac{e^{\ell_i}}
{\sum_j e^{\ell_j}}.
$$

These probabilities are combined into the signed interaction weight

$$
w
=
(p_{\text{agree}}-p_{\text{disagree}})
(1-p_{\text{neutral}}).
$$

The first factor determines stance direction and decisiveness. The second reduces the magnitude when the model assigns probability to the neutral class, shrinking ambiguous interactions towards zero while retaining larger magnitudes for clearer agreement or disagreement.

\subsection{Validation protocol and overlap with the analyzed corpus}

DEBAGREEMENT is sorted chronologically by post timestamp and divided into training, validation, and test partitions using an 80/10/10 split. The training period ends on 6 December 2020 and the validation period on 13 February 2021. The validation partition is used only for model selection, and the final 10\% is held out for classifier evaluation.

The training period contains approximately 87\% of the Brexit interactions used in the structural analysis. It contains all analyzed interactions from January 2019 through November 2020 and part of December 2020; the validation partition contains the remainder of December 2020 and January 2021. None of the analyzed months lies in the held-out test period.

The continuous network construction should therefore not be interpreted as statistically independent of the stance-model training data. The held-out results below evaluate classification on unseen interactions, but not the monthly Brexit networks themselves. The human-label construction provides an LLM-free structural reference.

\subsection{Held-out classification performance}

On the held-out test partition, the LoRA-adapted classifier achieves 84.5\% accuracy and a weighted F1 score of 0.846 across the three stance classes.

Because the downstream pipeline uses the continuous weight $w$ rather than only the top predicted label, uncertain interactions are attenuated rather than entering the structural analysis at the same magnitude as high-confidence predictions.

% =========================================================
\section{Synthetic benchmark details}
\label{sec:appendix_synthetic_generators}

This appendix provides the generator specifications and additional validation analyses for the five candidate polarization measures evaluated in Section~\ref{sec:benchmark_validation}. Their definitions are given in Section~\ref{sec:structural_polarization}. All measures are evaluated on the same generated graph instances so that observed differences arise from the measures rather than different graph substrates.

\subsection{Synthetic generators}

\paragraph{Signed stochastic block model.}
We generate signed stochastic block models with two equal-sized planted groups and $n=80$ nodes. Mean degree takes values

$$
d\in\{2,3,4,6,12,24\}.
$$

Within-group edges are assigned positive signs and between-group edges negative signs before independent sign noise is introduced. Each edge sign is flipped with probability

$$
\eta\in\{0,0.1,0.2,0.3,0.4,0.5\}.
$$

Thus $\eta=0$ represents a perfectly balanced planted two-camp structure, while $\eta=0.5$ removes the planted sign signal. We generate 30 replicates at mean degree $d\le3$ and 10 replicates at the remaining densities, giving 600 graphs.

\paragraph{Signed Barabási--Albert model.}
To test degree-heterogeneous networks and continuous edge magnitudes, we generate preferential-attachment graphs with attachment parameter

$$
m_{\mathrm{attach}}\in\{1,2,4,6\}.
$$

The $m_{\mathrm{attach}}=1$ setting produces trees and is included as a deliberately degenerate case because a forest contains no frustrated cycles.

Nodes are assigned latent binary labels. Edge weights are sampled from Gaussian distributions centred at $+1$ for same-group edges and $-1$ for cross-group edges, with standard deviation $0.3$. The same sign-noise grid $\eta\in\{0,0.1,\ldots,0.5\}$ is then applied.

We generate 30 replicates for $m_{\mathrm{attach}}=1$ and 10 replicates for each remaining attachment setting. Each graph is evaluated twice: once with its original continuous weights and once using a sign-only copy of the same graph.

\paragraph{Size sweep.}
To isolate sensitivity to graph size from density, we additionally generate signed block models with

$$
n\in\{80,160,320,640\},
$$

fixed mean degree $d\in\{3,6\}$ and
$\eta\in\{0.1,0.3,0.5\}$.
Ten replicates are generated for every setting.

All methods are run blind to the planted labels. Planted labels are used only for evaluation of partition recovery. Synthetic frustration optimization uses 800 random restarts followed by basin-hopping refinement.

\subsection{Sensitivity to planted sign structure}

We first test whether each score decreases as sign noise $\eta$ destroys the planted two-camp structure. Table~\ref{tab:synthetic_sensitivity} reports Spearman correlation between each score and $\eta$. Values before the slash use the complete grid; values after the slash exclude the sparse degenerate slice (degree 2 in the SBM and trees in the BA generator).

\begin{table*}[h]
\centering
\small
\begin{tabular}{lccc}
\toprule
Measure & SBM & BA continuous & BA sign-only \\
\midrule
$P^{\mathrm{eig}}$  & $-0.47/-0.75$ & $-0.33/-0.85$ & $-0.36/-0.88$ \\
$P^{\mathrm{gen}}$  & $-0.51/-0.59$ & $-0.30/-0.74$ & $-0.27/-0.76$ \\
$P^{\mathrm{edge}}$ & $-0.60/-0.73$ & $-0.34/-0.82$ & $-0.34/-0.82$ \\
$P^{\mathrm{c}}$    & $-0.77/-0.85$ & $-0.92/-0.92$ & $-0.92/-0.92$ \\
$P^{\mathrm{w}}$    & $-0.77/-0.85$ & $-0.91/-0.91$ & $-0.92/-0.92$ \\
\bottomrule
\end{tabular}
\caption{Sensitivity of each score to planted sign noise. Entries are Spearman $\rho$ between the score and $\eta$, for the complete grid and after excluding the sparse degenerate slice.}
\label{tab:synthetic_sensitivity}
\end{table*}

At $\eta=0$, the generalised spectral score, edge-normalized frustration score, and both cycle-normalized frustration scores equal $1$ in every density setting. The normalized Eigen-Sign score does not share this calibration: even on perfectly balanced graphs its value is approximately $0.014$, reflecting its size-dependent scale.

The cycle-normalized frustration measures are the most responsive to planted sign noise. Their sensitivity is weaker in very sparse graphs: for $P^{\mathrm{c}}$, the correlation with $\eta$ ranges from approximately $-0.61$ at degree 2 to $-0.96$ at degree 24.

\subsection{Specificity to density and graph size}

At $\eta=0.5$, the planted sign structure is absent, so systematic score variation across density is an artefact of the graph substrate rather than a response to polarization.

Across the synthetic density range, the cycle-normalized frustration scores vary by only $0.087$, compared with $0.607$ for edge-normalized frustration and $0.598$ for the generalised spectral score.

The Eigen-Sign score has a small absolute spread only because its complete numerical scale is compressed by graph size. In the independent size sweep, it correlates with $n$ at

$$
\rho=-0.965.
$$

The remaining measures show substantially smaller size dependence.

For the generalised spectral score, the $\eta=0.5$ values closely follow the random-sign spectral baseline

$$
\frac{2\sqrt{d-1}}{d}.
$$

At the mean-degree range observed in the Reddit networks, approximately $2.1$--$3.1$, this baseline is already about $0.91$--$0.98$ on a scale with ceiling $1$. This leaves little numerical headroom for additional signed structure.

Within fixed levels of $\eta$, residual density dependence of the cycle-normalized frustration scores varies in sign and largely cancels across the full grid. By contrast, the normalized, generalised Eigen-Sign, and edge-normalized frustration scores retain strong negative density relationships across noise levels.

\subsection{Partition recovery and agreement}

We evaluate the partitions recovered by the generalised spectral method and weighted frustration against the planted labels using Adjusted Rand Index (ARI).

At low noise, both methods recover the planted partition similarly. For example, at $\eta=0.1$, their respective ARI values are approximately $0.61/0.64$ at degree 3, $0.95/0.93$ at degree 6, and $0.99/1.00$ at degree 12. Both methods lose planted-label recovery as the signal approaches the detectability limit.

Their mutual partition agreement on the signed SBM is shown in Table~\ref{tab:synthetic_ari_sbm}.

\begin{table*}[h]
\centering
\small
\begin{tabular}{lrrrrrr}
\toprule
Mean degree & $\eta=0$ & $0.1$ & $0.2$ & $0.3$ & $0.4$ & $0.5$ \\
\midrule
2  & 1.00 & 0.72 & 0.66 & 0.58 & 0.62 & 0.70 \\
3  & 1.00 & 0.76 & 0.41 & 0.41 & 0.45 & 0.46 \\
6  & 1.00 & 0.95 & 0.69 & 0.34 & 0.35 & 0.33 \\
12 & 1.00 & 0.99 & 0.96 & 0.62 & 0.41 & 0.32 \\
24 & 1.00 & 1.00 & 0.99 & 0.94 & 0.34 & 0.35 \\
\bottomrule
\end{tabular}
\caption{Mean partition agreement, measured by ARI, between the blind generalised spectral and weighted-frustration partitions on the signed SBM.}
\label{tab:synthetic_ari_sbm}
\end{table*}

Table~\ref{tab:synthetic_ari_ba} reports the same comparison for the continuously weighted Barabási--Albert graphs.

\begin{table}[h]
\centering
\small
\begin{tabular}{lrrrrrr}
\toprule
$m_{\mathrm{attach}}$ & $\eta=0$ & $0.1$ & $0.2$ & $0.3$ & $0.4$ & $0.5$ \\
\midrule
1 & 1.00 & 1.00 & 1.00 & 1.00 & 1.00 & 1.00 \\
2 & 1.00 & 0.83 & 0.59 & 0.44 & 0.41 & 0.34 \\
4 & 1.00 & 0.97 & 0.79 & 0.53 & 0.31 & 0.41 \\
6 & 1.00 & 0.98 & 0.95 & 0.67 & 0.41 & 0.50 \\
\bottomrule
\end{tabular}
\caption{Mean partition agreement between the generalised spectral and weighted-frustration partitions on continuously weighted signed Barabási--Albert graphs. The tree row is structurally degenerate because forests contain no frustrated cycles.}
\label{tab:synthetic_ari_ba}
\end{table}

Although the two methods agree strongly on partitions when a planted signal is recoverable, their score-level agreement is largely induced by their shared response to $\eta$. Overall Spearman correlation between the two scores is $+0.52$ on the SBM and $+0.70$ on the continuously weighted BA graphs. Within a fixed noise level, the mean correlation falls to approximately $+0.05$ and $0$, respectively.

\subsection{Spectral pedestal diagnostic}

To determine whether the high generalised spectral score on random-sign sparse graphs is caused by the normalization or by maximisation over possible assignments, we evaluate the same generalised Rayleigh quotient using the known planted assignment rather than the maximising eigenvector.

At $\eta=0.5$, the oracle value remains near zero at every density, approximately $-0.03$ to $+0.01$, whereas the maximised score ranges from approximately $0.38$ at degree 24 to $0.98$ at degree 2. The density-dependent pedestal therefore arises from maximisation over random sign structure rather than from the normalizing denominator itself.

\subsection{Continuous versus sign-only frustration}

On paired Barabási--Albert graphs, weighted frustration applied to continuous weights and count frustration applied to sign-only copies behave similarly under planted noise. Their sensitivities are $-0.910$ and $-0.917$, their score rankings correlate at $0.975$, and the mean absolute difference between their scores is $0.022$. Their planted-partition recovery differs by less than $0.003$ ARI.

This comparison shows that the weighted formulation retains edge-magnitude information without losing the structural behaviour of the count formulation.

\subsection{Summary of synthetic validation}

Table~\ref{tab:synthetic_decision} summarises the synthetic validation criteria on the SBM benchmark.

\begin{table*}[h]
\centering
\small
\setlength{\tabcolsep}{4pt}
\begin{tabular}{lrrrrrr}
\toprule
Measure &
Sensitivity $\rho(\eta)$ &
Density $\rho$ &
Size $\rho$ &
$\eta=0.5$ spread &
$\eta=0$ value &
Recovery ARI \\
\midrule
$P^{\mathrm{eig}}$  & $-0.47$ & $-0.74$ & $-0.97$ & $0.011$ & $0.014$ & $0.19$ \\
$P^{\mathrm{gen}}$  & $-0.51$ & $-0.65$ & $+0.14$ & $0.598$ & $1.000$ & $0.17$ \\
$P^{\mathrm{edge}}$ & $-0.60$ & $-0.60$ & $-0.09$ & $0.607$ & $1.000$ & $0.21$ \\
$P^{\mathrm{c}}$    & $-0.77$ & $+0.01$ & $-0.14$ & $0.087$ & $1.000$ & $0.21$ \\
$P^{\mathrm{w}}$    & $-0.77$ & $+0.01$ & $-0.14$ & $0.087$ & $1.000$ & $0.21$ \\
\bottomrule
\end{tabular}
\caption{Summary of synthetic score validity. Recovery ARI is reported at $\eta=0.2$ and mean degree 3.}
\label{tab:synthetic_decision}
\end{table*}

% =========================================================
\section{Real-network validation details}
\label{sec:appendix_real_sanity}

This appendix reports additional validation results for the monthly Reddit Brexit networks. The primary analyses use the 24-month sample defined in Section~\ref{sec:sample_rule}. Covariate structure is additionally examined on all 30 graphs because size and density remain defined for acyclic windows.

\subsection{Covariate structure}

Graph size and density characteristics are strongly correlated with one another. On the continuous largest-component graphs across all 30 windows,

$$
\rho(n,\text{density})=-0.968,
$$

while

$$
\rho(n,m)=+0.985
$$

and

$$
\rho(n,\text{mean degree})=+0.818.
$$

Similarly,

$$
\rho(m,\text{density})=-0.929
$$

and

$$
\rho(m,\text{mean degree})=+0.883.
$$

Across node count, edge count, mean degree, density, cyclomatic number, total absolute weight, and weighted cycle capacity, pairwise absolute correlations lie between approximately $0.68$ and $0.99$.

Because these graph characteristics are highly collinear, we evaluate confounding with a family-wise test based on the maximum absolute correlation across node count, edge count, mean degree, and cyclomatic number rather than separate unadjusted tests for each variable.

\subsection{Size and density screen}

Table~\ref{tab:screen} reports Spearman correlation between each candidate polarization score and the four graph characteristics on the 24 analyzed continuous-construction windows. The family-wise $p$-value is obtained from 5{,}000 shared permutations of the maximum absolute correlation statistic.

\begin{table*}[h]
\centering
\small
\setlength{\tabcolsep}{4pt}
\begin{tabular}{lrrrrrr}
\toprule
Score & vs $n$ & vs $m$ & vs deg. & vs $\kappa$ & $\max|\rho|$ & $p_{\mathrm{fw}}$ \\
\midrule
$P^{\mathrm{eig}}$  & $-0.959$ & $-0.989$ & $-0.798$ & $-0.923$ & $0.989$ & $<0.001$ \\
$P^{\mathrm{gen}}$  & $-0.214$ & $-0.219$ & $-0.387$ & $-0.323$ & $0.387$ & $0.115$ \\
$P^{\mathrm{edge}}$ & $-0.627$ & $-0.716$ & $-0.941$ & $-0.872$ & $0.941$ & $<0.001$ \\
$P^{\mathrm{c}}$    & $-0.413$ & $-0.386$ & $-0.456$ & $-0.432$ & $0.456$ & $0.0498$ \\
$P^{\mathrm{w}}$    & $+0.246$ & $+0.187$ & $+0.049$ & $+0.077$ & $0.246$ & $0.394$ \\
\bottomrule
\end{tabular}
\caption{Size and density screen on the 24 analyzed months using the continuous construction and largest connected component. $p_{\mathrm{fw}}$ is the family-wise permutation $p$-value of the maximum absolute Spearman correlation.}
\label{tab:screen}
\end{table*}

The normalized Eigen-Sign score is strongly associated with node and edge counts and is therefore close to a readout of window activity. The edge-normalized frustration score is similarly strongly associated with mean degree.

The weighted cycle-normalized frustration score shows no substantial association with the screened graph characteristics. The count version lies on the boundary of the family-wise criterion because of its negative correlation with mean degree.

On the human-label construction, where $P^{\mathrm{c}}=P^{\mathrm{w}}$, the maximum absolute correlation is $0.297$ with $p_{\mathrm{fw}}=0.27$. By contrast, the spectral and edge-normalized baselines fail strongly on the human-labelled graphs.

\subsection{Substrate robustness}

We repeat the size/density screen on each graph's 2-core. The weighted frustration index again passes, with maximum absolute correlation $0.172$ and $p=0.67$. The count index gives $0.428$ with $p=0.080$.

The generalised spectral score moves in the opposite direction: it passes the largest-component screen but fails on the 2-core, where its maximum absolute correlation reaches $0.596$ with $p=0.007$.

Month-ranking stability between the largest connected component and the 2-core is

$$
\rho=1.000
$$

for $P^{\mathrm{w}}$,

$$
\rho=0.978
$$

for $P^{\mathrm{c}}$,

$$
\rho=0.973
$$

for $P^{\mathrm{eig}}$,

$$
\rho=0.919
$$

for $P^{\mathrm{edge}}$, and only

$$
\rho=0.383
$$

for $P^{\mathrm{gen}}$.

Leave-one-window-out values of the weighted index's maximum size/density correlation remain between $0.15$ and $0.35$, with standard deviation $0.05$, indicating that no single month drives its validation result.

\subsection{Reliability and invariance}

Test--retest reliability is assessed by randomly retaining 95\% of edges in each window over 40 draws and recomputing the scores at a reduced 1{,}500-restart optimization budget.

On the continuous construction, the resulting intraclass correlations are

\begin{table}[h]
\centering
\small
\begin{tabular}{lc}
\toprule
Score & ICC \\
\midrule
$P^{\mathrm{eig}}$  & 0.99 \\
$P^{\mathrm{edge}}$ & 0.95 \\
$P^{\mathrm{gen}}$  & 0.79 \\
$P^{\mathrm{c}}$    & 0.71 \\
$P^{\mathrm{w}}$    & 0.70 \\
\bottomrule
\end{tabular}
\caption{Test--retest ICC under 95\% edge subsampling on the continuous construction.}
\label{tab:reliability}
\end{table}

The very high reliability of $P^{\mathrm{eig}}$ and $P^{\mathrm{edge}}$ should not be interpreted as evidence of validity: those two scores are strongly coupled to graph size, which changes little under small edge perturbations.

On the human-label construction, $P^{\mathrm{c}}=P^{\mathrm{w}}$ gives ICC $0.79$, while the generalised spectral score falls to $0.37$.

Additional checks confirm deterministic reruns, invariance of the weighted scores to multiplying all magnitudes by a common positive constant, and the bounds

$$
F_t\le\kappa_t
$$

and

$$
F^{\mathrm{w}}_t\le C^{\mathrm{w}}_t
$$

in every analyzed window.

\subsection{Agreement between measures}

The generalised spectral and weighted-frustration scores show little cross-window agreement on the 24 real months:

$$
r=0.01,
\qquad
\rho=-0.09.
$$

Within families, the spectral measures correlate at $\rho=0.25$. The edge-normalized frustration score correlates with weighted frustration at $\rho=0.03$ but with count frustration at $\rho=0.67$.

The two frustration measures carried into the language analysis correlate at

$$
r=0.52,
\qquad
\rho=0.20,
$$

showing that they provide related but non-interchangeable views of the monthly graphs.

Against the human-label structural score, weighted frustration has rank correlation

$$
\rho=+0.64,
$$

compared with

$$
\rho=+0.31
$$

for count frustration.

\subsection{optimization convergence}

Plain multistart descent continues to improve between 10{,}000 and 20{,}000 restarts in 22 of 90 real-data window/objective combinations. Basin-hopping refinement improves on the plain 20{,}000-restart solution in 40 of 90 combinations.

Two fully independent full-budget searches agree in 84 of the 90 cases. All six disagreements occur for the count objective in the largest windows and differ by only one frustrated edge. The weighted objective agrees exactly in all 30 monthly windows.

All reported frustration values should therefore be interpreted as numerical upper bounds on the global optimum, with slightly greater residual optimization uncertainty for the count measure.

% =========================================================
\section{Sample rule and sign-shuffle null}
\label{sec:appendix_sample_rule}

\subsection{Sample-rule rationale}

The cycle-normalized frustration indices are defined only when the graph contains at least one independent cycle. Moreover, when only one or two cycles are present, the score has very few attainable values and can be dominated by individual cycle constraints.

We therefore use one common sample rule for all real-data analyses: the continuous graph must have a non-empty cycle space, and the human-label graph must contain at least three independent cycles.

To examine score resolution directly, we generate 200 random sign assignments on each observed graph while holding its topology fixed. The standard deviation of the resulting frustration index decreases approximately as

$$
\frac{0.5}{\sqrt{k}},
$$

where $k$ is the number of independent cycles. A one-cycle graph therefore behaves approximately as a balanced-or-unbalanced coin flip, motivating the minimum-cycle requirement.

Of the 30 monthly windows, 24 satisfy the common rule. They form a series from January 2019 to January 2021 with one interior gap, November 2019.

\begin{table}[h]
\centering
\small
\begin{tabular}{ll}
\toprule
Window & Reason for exclusion \\
\midrule
2018-12 & one human-graph cycle \\
2019-11 & forest under human labels \\
2021-02 & forest under human labels \\
2021-03 & forest under both constructions \\
2021-04 & one human-graph cycle \\
2021-05 & forest under both constructions \\
\bottomrule
\end{tabular}
\caption{Windows excluded by the common sample rule. Five of the six exclusions occur at the beginning or end of the observation period; November 2019 is the only interior exclusion.}
\label{tab:excluded}
\end{table}

June 2019 is retained but contains exactly three independent cycles in the human-label graph, all of which are balanced. It is therefore treated as a leverage check throughout the language analysis.

\subsection{Per-window sign-shuffle null}

The size/density screen tests whether a score varies across months with graph characteristics that are not signed structure. We additionally ask a different question: whether the observed sign organisation within an individual month is distinguishable from random placement of the same signs on the same graph.

For each largest-component graph, we retain the topology and every edge magnitude while permuting the signs across edges. This preserves the number of positive and negative edges and the distribution of absolute weights, while destroying the correspondence between edge location and sign. We generate 200 replicates per window and graph construction.

Observed and shuffled scores are computed using the same reduced optimization budget. For each window we record the null mean, null standard deviation, and percentile of the observed score.

On the 24 analyzed months, the weighted frustration score exceeds its own sign-shuffle null mean in 15 months and falls below it in nine. Nineteen of the 24 observations lie within one null standard deviation. Only two months exceed the 95th percentile of their weighted null distribution.

For the count score, four months exceed the 95th percentile. These results indicate that the empirical series should primarily be interpreted comparatively across months rather than as evidence that every individual month displays signed organisation beyond its own random-sign baseline.

The months above the 90th percentile of the weighted-frustration null are shown in Table~\ref{tab:sign_shuffle_top}.

\begin{table}[h]
\centering
\small
\begin{tabular}{lrrrr}
\toprule
Month & $\kappa$ & Observed & Null mean $\pm$ sd & Percentile \\
\midrule
2020-11 & 21 & 0.862 & $0.588\pm0.117$ & 0.965 \\
2020-05 & 34 & 0.707 & $0.608\pm0.062$ & 0.950 \\
2019-12 & 42 & 0.729 & $0.657\pm0.050$ & 0.915 \\
\bottomrule
\end{tabular}
\caption{Months above the 90th percentile of their weighted-frustration sign-shuffle null.}
\label{tab:sign_shuffle_top}
\end{table}

% =========================================================
\section{Scalar-intensity lexicon generation and robustness}
\label{sec:scalar_lexicon}

\subsection{Lexicon generation}

To reduce dependence on a single generated word list, we generate the scalar-intensity lexicon five times using one identical GPT-5 prompt. The five runs differ only through model sampling.

The prompt defines scalar intensity broadly to include intensifiers, evaluative language, emotion terms, Brexit-specific stance terms, and politically charged descriptors. Ten fixed calibration anchors establish a common score scale across generations.

The complete prompt is:

\begin{quote}
\small
\textit{System:}
``You are a linguistic expert analyzing political discourse. Return a JSON object with a 'lexicon' key containing a list of \{term, score\} objects. Terms are lowercase single words; scores are floats between 0 and 1 measuring the term's scalar intensity. Calibrate scores to this fixed scale: somewhat = 0.3, fairly = 0.4, pretty = 0.5, quite = 0.6, very = 0.7, really = 0.75, extremely = 0.85, incredibly = 0.9, absolutely = 0.95, completely = 1.0. Provide about 100 entries.''

\medskip

\textit{User:}
``You are curating a scalar-intensity lexicon for Reddit discussions of Brexit. Cover, in one flat list: intensifier adverbs; common evaluative adjectives, both positive (e.g. from small improvements up to perfect) and negative (from minor up to horrendous); emotion words from mild concern up to extreme fear or anger; Brexit-specific actor and stance terms (like remainer, brexiteer, eurosceptic); and politically charged descriptors (like biased, corrupt, xenophobic, dishonest). Spread the scores across the whole calibrated range from mild (about 0.3) to extreme (0.9 and above), with at least 10 terms scoring 0.9 or higher. Return JSON mapping words to [0,1].''
\end{quote}

The five generations contain 103, 139, 100, 121, and 100 terms, respectively. The corresponding numbers of terms scoring at least $0.9$ are 23, 23, 23, 19, and 12. Forty-four terms occur in all five lexicons, pairwise Jaccard overlap ranges from $0.41$ to $0.51$, and all ten calibration anchors are reproduced at the requested scores in every generation.

No hand-written list is blended into the generated lexicons.

\subsection{Feature construction}

For each generated lexicon, every comment–reply pair is assigned the maximum intensity score of any matched term, or $0$ when no term is present.

From these per comment–reply pairs maximum we compute four thresholded shares:

$$
x^{\mathrm{any}}_t:
\quad
\text{maximum}>0,
$$

$$
x^{0.5}_t:
\quad
\text{maximum}\ge0.5,
$$

$$
x^{0.7}_t:
\quad
\text{maximum}\ge0.7,
$$

and

$$
x^{\mathrm{scalar}}_t:
\quad
\text{maximum}\ge0.9.
$$

We additionally compute the mean per-post maximum,
$x^{\mathrm{mean}}_t$.

The main analysis uses the mean of each feature across the five generated lexicons.

Across the 24 analyzed months, the resulting ranges are approximately:

\begin{table}[h]
\centering
\small
\begin{tabular}{lc}
\toprule
Feature & Range across months \\
\midrule
Any scalar term & $0.538$--$0.716$ \\
Maximum $\ge0.5$ & $0.493$--$0.672$ \\
Maximum $\ge0.7$ & $0.217$--$0.305$ \\
Maximum $\ge0.9$ & $0.054$--$0.108$ \\
Mean maximum intensity & $0.348$--$0.446$ \\
\bottomrule
\end{tabular}
\caption{Ranges of the scalar-intensity features on the 24 analyzed months.}
\label{tab:scalar_ranges}
\end{table}

\subsection{Agreement across lexicon generations}

Because the generated lexicons differ in size, their absolute feature levels differ. Their ranking of months is nevertheless substantially more stable.

Pairwise correlations between the monthly $x^{\mathrm{any}}_t$ series range from $+0.64$ to $+0.96$, with mean $+0.80$. For mean scalar intensity they range from $+0.71$ to $+0.92$. The extreme-tail feature is less reproducible, with pairwise correlations from $+0.54$ to $+0.93$.

\subsection{Run-level association robustness}

Table~\ref{tab:lexicon_runs} reports Pearson correlation between selected scalar features and the count, weighted, and human-label polarization scores separately for each generated lexicon. Circular-shift $p$-values are shown in parentheses.

\begin{table*}[h]
\centering
\small
\setlength{\tabcolsep}{5pt}
\begin{tabular}{llccc}
\toprule
Feature & Run & $P^{\mathrm{c}}$ & $P^{\mathrm{w}}$ & $P^{\mathrm{H}}$ \\
\midrule
$x^{\mathrm{any}}$
& 0 & $+0.56\;(0.042)$ & $+0.39\;(0.13)$ & $+0.36\;(0.13)$ \\
& 1 & $+0.61\;(0.042)$ & $+0.26\;(0.33)$ & $+0.40\;(0.083)$ \\
& 2 & $+0.50\;(0.083)$ & $+0.46\;(0.083)$ & $+0.40\;(0.13)$ \\
& 3 & $+0.35\;(0.17)$  & $+0.27\;(0.21)$ & $+0.37\;(0.042)$ \\
& 4 & $+0.55\;(0.042)$ & $+0.29\;(0.21)$ & $+0.27\;(0.25)$ \\
& mean & $+0.56\;(0.042)$ & $+0.36\;(0.13)$ & $+0.39\;(0.13)$ \\
\midrule
$x^{\mathrm{mean}}$
& 0 & $+0.50\;(0.13)$ & $+0.43\;(0.083)$ & $+0.35\;(0.13)$ \\
& 1 & $+0.56\;(0.042)$ & $+0.27\;(0.29)$ & $+0.36\;(0.083)$ \\
& 2 & $+0.41\;(0.17)$ & $+0.40\;(0.083)$ & $+0.24\;(0.21)$ \\
& 3 & $+0.39\;(0.13)$ & $+0.30\;(0.21)$ & $+0.40\;(0.042)$ \\
& 4 & $+0.51\;(0.042)$ & $+0.31\;(0.25)$ & $+0.25\;(0.17)$ \\
& mean & $+0.52\;(0.083)$ & $+0.37\;(0.13)$ & $+0.35\;(0.13)$ \\
\midrule
$x^{\mathrm{scalar}}$
& 0 & $-0.09\;(0.63)$ & $+0.06\;(1.00)$ & $-0.01\;(1.00)$ \\
& 1 & $-0.06\;(0.83)$ & $+0.10\;(0.79)$ & $+0.02\;(0.96)$ \\
& 2 & $-0.15\;(0.38)$ & $-0.19\;(0.38)$ & $-0.29\;(0.29)$ \\
& 3 & $-0.11\;(0.63)$ & $+0.05\;(0.92)$ & $-0.02\;(0.96)$ \\
& 4 & $+0.09\;(0.71)$ & $+0.20\;(0.38)$ & $+0.16\;(0.46)$ \\
& mean & $-0.08\;(0.79)$ & $+0.04\;(0.96)$ & $-0.04\;(0.79)$ \\
\bottomrule
\end{tabular}
\caption{Run-level scalar-language associations. Entries are Pearson $r$ with circular-shift $p$-values in parentheses.}
\label{tab:lexicon_runs}
\end{table*}

The broad prevalence association with the count score is positive in all five runs, ranging from $+0.35$ to $+0.61$. By contrast, the extreme-tail feature changes sign across lexicon generations and shows no stable association with any structural score. This motivates treating scalar prevalence rather than the extreme tail as the primary language feature.

% =========================================================
\section{Association adjustment and prediction details}
\label{sec:appendix_adjustment}

This appendix provides the full association results and robustness analyses for Section~\ref{sec:linking_polarization_to_language}, including the language-feature screen, the human-label reference, covariate adjustment, leverage analysis, and the forecasting and nowcasting experiments.

% =========================================================
\subsection{Full contemporaneous association results}

Table~\ref{tab:association_full} reports the contemporaneous associations between the nine language features and the count and weighted frustration indices over the 24 analyzed months. For each feature, we report Pearson $r$, Spearman $\rho$, and the circular-shift $p$-value.

\begin{table*}[t]
\centering
\small
\setlength{\tabcolsep}{4pt}
\begin{tabular}{lrrrrrr}
\toprule
& \multicolumn{3}{c}{Count $P^{\mathrm{c}}$}
& \multicolumn{3}{c}{Weighted $P^{\mathrm{w}}$} \\
\cmidrule(lr){2-4}\cmidrule(lr){5-7}
Feature
& $r$ & $\rho$ & $p_{\mathrm{shift}}$
& $r$ & $\rho$ & $p_{\mathrm{shift}}$ \\
\midrule
Prevalence $x^{\mathrm{any}}$
& $+0.561$ & $+0.539$ & $0.042$
& $+0.363$ & $+0.325$ & $0.125$ \\

Share $\ge0.5$
& $+0.566$ & $+0.553$ & $0.042$
& $+0.358$ & $+0.296$ & $0.167$ \\

Share $\ge0.7$
& $-0.052$ & $0.000$ & $0.83$
& $+0.154$ & $+0.130$ & $0.63$ \\

Extreme share $x^{\mathrm{scalar}}$
& $-0.078$ & $+0.180$ & $0.79$
& $+0.040$ & $+0.120$ & $0.96$ \\

Mean intensity $x^{\mathrm{mean}}$
& $+0.516$ & $+0.530$ & $0.083$
& $+0.370$ & $+0.370$ & $0.125$ \\

Extreme toxicity $x^{\mathrm{tox}}$
& $-0.168$ & $-0.380$ & $0.50$
& $+0.109$ & $+0.190$ & $0.71$ \\

Mean toxicity
& $+0.165$ & $-0.030$ & $0.42$
& $+0.156$ & $+0.240$ & $0.54$ \\

Maximum toxicity
& $+0.005$ & $-0.190$ & $1.00$
& $-0.003$ & $+0.320$ & $0.96$ \\

Mean perplexity
& $-0.173$ & $-0.210$ & $0.54$
& $-0.156$ & $-0.170$ & $0.42$ \\
\bottomrule
\end{tabular}
\caption{Contemporaneous associations between the nine language features and the count and weighted frustration indices over the 24 analyzed months.}
\label{tab:association_full}
\end{table*}

The broad scalar-intensity features show the clearest relationship with the count index. Scalar prevalence and the share above $0.5$ both exceed every circular shift, while the relationship weakens at higher thresholds and is absent in the extreme tail. Toxicity and perplexity show no comparable relationship with either frustration measure.

For Pearson correlations, scalar prevalence and the share above $0.5$ both have Holm-adjusted $p=0.036$ and Benjamini--Hochberg $q=0.020$ on the count index. Mean scalar intensity has Holm-adjusted $p=0.069$ and $q=0.030$. No feature survives Holm correction on the weighted index. Because the minimum circular-shift value is $1/24$, shift-based $p$-values cannot survive a nine-feature Holm correction.

% =========================================================
\subsection{Human-label reference}

We repeat the association analysis using the count frustration index computed directly from the human agreement, disagreement, and neutral annotations. This construction does not use the stance model and provides an LLM-free reference.

\begin{table}[t]
\centering
\small
\setlength{\tabcolsep}{4pt}
\begin{tabular}{lrrrr}
\toprule
Feature & $r$ & $\rho$ & $p(\rho)$ & $p_{\mathrm{shift}}$ \\
\midrule
Prevalence $x^{\mathrm{any}}$
& $+0.39$ & $+0.48$ & $0.017$ & $0.125$ \\

Mean intensity $x^{\mathrm{mean}}$
& $+0.35$ & $+0.47$ & $0.020$ & $0.125$ \\

Share $\ge0.5$
& $+0.31$ & $+0.38$ & $0.066$ & $0.125$ \\

Extreme share $x^{\mathrm{scalar}}$
& $-0.04$ & $+0.09$ & $0.68$ & $0.79$ \\

Extreme toxicity $x^{\mathrm{tox}}$
& $+0.35$ & $+0.32$ & $0.13$ & $0.083$ \\

Mean toxicity
& $+0.39$ & $+0.37$ & $0.075$ & $0.083$ \\

Maximum toxicity
& $+0.16$ & $+0.46^{\ddagger}$ & $0.024$ & $0.50$ \\

Mean perplexity
& $+0.12$ & $+0.02$ & $0.94$ & $0.67$ \\
\bottomrule
\end{tabular}
\caption{Associations between the language features and the human-label frustration index. $^{\ddagger}$Maximum toxicity is ceiling-saturated in most months, so its rank correlation is not treated as a stable association.}
\label{tab:association_human}
\end{table}

No feature exceeds the circular-shift null on the human-label reference. Scalar prevalence and mean scalar intensity nevertheless show positive rank associations similar in direction to those observed for the count index. We therefore treat the human-label results as supporting rather than confirmatory evidence.

The threshold and lexicon-regeneration analyses are reported separately in Appendix~\ref{sec:scalar_lexicon}.

% =========================================================
\subsection{Leverage analysis}

June 2019 is a leverage point in the contemporaneous association. It has the highest scalar prevalence in the retained sample and a high count frustration score. Removing this month reduces the scalar-prevalence association with the count index from

$$
r=+0.56
$$

to

$$
r=+0.39,
$$

with $p=0.066$. The corresponding rank association remains

$$
\rho=+0.48,
$$

with $p=0.022$.

For the weighted index, the raw Pearson association decreases from $+0.36$ to $+0.15$ when June 2019 is removed. These results indicate that the magnitude of the contemporaneous relationship is sensitive to this month, although the positive rank relationship for the count index remains.

% =========================================================
\subsection{Covariate adjustment}

We test whether the scalar-prevalence relationship is explained by the other language dimensions or by variation in graph size and density. All predictors are standardised before fitting ordinary least-squares models.

The \emph{language-adjusted} model includes scalar prevalence, mean toxicity, and mean perplexity. The \emph{graph-adjusted} model includes scalar prevalence, node count, and mean degree. The \emph{combined} model includes both adjustment blocks.

For each specification, we report the standardised coefficient of scalar prevalence, its $t$-statistic and parametric $p$-value, its partial correlation with the structural score, and a conditional circular-shift $p$-value. In the conditional shift test, scalar prevalence alone is rotated while the adjustment variables remain fixed.

\subsubsection{Count frustration index}

\begin{table*}[t]
\centering
\small
\setlength{\tabcolsep}{4pt}
\begin{tabular}{llrrrrrrr}
\toprule
Sample & Adjustment & $\beta$ & $t$ & $p$ &
Partial $r$ & Shift $p$ & $R^2$ & LOO $R^2$ \\
\midrule
All 24
& Language
& $+0.027$ & $3.02$ & $0.007$
& $+0.56$ & $0.042$ & $0.36$ & $-0.13$ \\

All 24
& Graph
& $+0.019$ & $2.61$ & $0.017$
& $+0.50$ & $0.125$ & $0.58$ & $+0.30$ \\

All 24
& Both
& $+0.018$ & $2.48$ & $0.023$
& $+0.50$ & $0.083$ & $0.65$ & $+0.26$ \\

Without 2019-06
& Language
& $+0.013$ & $1.97$ & $0.063$
& $+0.41$ & $0.087$ & $0.27$ & $-0.08$ \\

Without 2019-06
& Graph
& $+0.010$ & $1.79$ & $0.089$
& $+0.38$ & $0.26$ & $0.42$ & $+0.18$ \\

Without 2019-06
& Both
& $+0.010$ & $1.89$ & $0.076$
& $+0.42$ & $0.13$ & $0.54$ & $+0.18$ \\
\bottomrule
\end{tabular}
\caption{Covariate-adjusted association between scalar prevalence and the count frustration index. LOO denotes leave-one-out $R^2$.}
\label{tab:adjustment_count}
\end{table*}

Adjustment for mean toxicity and mean perplexity leaves the scalar-prevalence association essentially unchanged. When node count and mean degree are included, the partial correlation remains similar in magnitude, but the conditional shift test no longer reaches its minimum value. Mean degree also contributes to variation in the count index. The association is therefore robust to the language covariates but less certain after adjustment for graph size and density.

Removing June 2019 weakens the scalar-prevalence term under every specification, although its direction remains positive.

\subsubsection{Weighted frustration index}

\begin{table*}[t]
\centering
\small
\setlength{\tabcolsep}{4pt}
\begin{tabular}{lrrrrrrr}
\toprule
Adjustment & $\beta$ & $t$ & $p$ &
Partial $r$ & Shift $p$ & $R^2$ & LOO $R^2$ \\
\midrule
Language
& $+0.030$ & $1.69$ & $0.107$
& $+0.35$ & $0.17$ & $0.17$ & $-0.33$ \\

Graph
& $+0.028$ & $1.56$ & $0.134$
& $+0.33$ & $0.17$ & $0.24$ & $-0.21$ \\

Both
& $+0.026$ & $1.41$ & $0.175$
& $+0.32$ & $0.29$ & $0.26$ & $-0.58$ \\
\bottomrule
\end{tabular}
\caption{Covariate-adjusted scalar-prevalence results for the weighted frustration index on the full 24-month sample.}
\label{tab:adjustment_weighted}
\end{table*}

The weighted association remains weaker under all three adjustment specifications. When June 2019 is removed, the partial correlations fall below $+0.18$ and none of the adjusted scalar-prevalence terms approaches significance.

These regressions are intended as adjustment checks for the contemporaneous association rather than as predictive models.

% =========================================================
\subsection{Prediction design}

The 24 retained months yield 22 consecutive-month pairs. January 2019 has no preceding month in the retained sample, and December 2019 cannot form a lagged pair because November 2019 is excluded.

The prediction experiments use Random Forest regression with 500 trees and a fixed random seed. The structure-only model uses the previous month's structural score,

$$
[P_{t-1}],
$$

while the augmented model additionally includes scalar prevalence. For one-month-ahead forecasting, the added feature is

$$
x^{\mathrm{any}}_{t-1},
$$

whereas contemporaneous nowcasting uses

$$
x^{\mathrm{any}}_{t}.
$$

Prediction gain is defined as

$$
\mathrm{Gain}
=
1-
\frac{\mathrm{MSE}_{\mathrm{aug}}}
     {\mathrm{MSE}_{\mathrm{base}}}.
$$

The primary evaluation uses rolling origin. After a 10-pair burn-in, each of the remaining 12 months is predicted using models fitted only to earlier pairs. A chronological 80/20 split with 18 training pairs and four test pairs is retained as a secondary comparison for the forecasting analysis.

For the permutation test, the language-feature sequence is rotated through all 21 non-identity offsets and the complete evaluation is repeated. The resulting value is

$$
p_{\mathrm{shift}}
=
\frac{
1+
\#\{
\mathrm{Gain}_{k}
\ge
\mathrm{Gain}_{\mathrm{obs}}
\}
}{22},
$$

with a minimum attainable value of

$$
\frac{1}{22}
\approx0.045.
$$

% =========================================================
\subsection{One-month-ahead forecasting}

Table~\ref{tab:forecast_appendix} compares the structure-only and augmented models when scalar prevalence is measured in the previous month.

\begin{table*}[t]
\centering
\small
\setlength{\tabcolsep}{4pt}
\begin{tabular}{llrrrrr}
\toprule
Target & Evaluation & Test months & MSE base & MSE aug. & Gain & $p_{\mathrm{shift}}$ \\
\midrule
Count $P^{\mathrm{c}}$
& 80/20 & 4
& $0.0050$ & $0.0063$ & $-26.0\%$ & $0.95$ \\

Count $P^{\mathrm{c}}$
& Rolling & 12
& $0.0033$ & $0.0031$ & $+8.4\%$ & $0.95$ \\

Weighted $P^{\mathrm{w}}$
& 80/20 & 4
& $0.0179$ & $0.0168$ & $+6.3\%$ & $0.55$ \\

Weighted $P^{\mathrm{w}}$
& Rolling & 12
& $0.0094$ & $0.0087$ & $+6.6\%$ & $0.36$ \\

Human $P^{\mathrm{H}}$
& 80/20 & 4
& $0.0043$ & $0.0012$ & $+71.1\%$ & $0.23$ \\

Human $P^{\mathrm{H}}$
& Rolling & 12
& $0.0112$ & $0.0114$ & $-2.1\%$ & $0.41$ \\
\bottomrule
\end{tabular}
\caption{One-month-ahead forecasting using previous-month scalar prevalence. The rolling-origin evaluation is the primary result; the chronological 80/20 split contains only four test months and is shown as a secondary comparison.}
\label{tab:forecast_appendix}
\end{table*}

No forecasting result exceeds its circular-shift null. The large gain for the human-label score under the four-month chronological split does not persist under rolling-origin evaluation. Replacing scalar prevalence with mean scalar intensity does not change this conclusion.

A linear lagged model gives the same result. In models of the form

$$
P_t
=
\beta_0
+
\beta_1 P_{t-1}
+
\beta_2 x^{\mathrm{any}}_{t-1}
+
\epsilon_t,
$$

the lagged scalar-prevalence term is null for the count, weighted, and human-label scores.

The monthly structural scores also show little persistence:

$$
r(P^{\mathrm{c}}_{t-1},P^{\mathrm{c}}_t)
=
+0.17,
$$

$$
r(P^{\mathrm{w}}_{t-1},P^{\mathrm{w}}_t)
=
+0.01,
$$

and

$$
r(P^{\mathrm{H}}_{t-1},P^{\mathrm{H}}_t)
=
-0.13.
$$

Scalar prevalence itself has weak monthly autocorrelation,

$$
r(x^{\mathrm{any}}_{t-1},x^{\mathrm{any}}_t)
=
+0.18.
$$

Together, these results provide no evidence that the previous month's scalar prevalence forecasts structural polarization one month later.

% =========================================================
\subsection{Contemporaneous nowcasting}

For nowcasting, the augmented model uses the current month's scalar prevalence together with the previous structural score. Table~\ref{tab:nowcast_appendix} reports the rolling-origin results.

\begin{table*}[t]
\centering
\small
\setlength{\tabcolsep}{4pt}
\begin{tabular}{llrrrrr}
\toprule
Target & Model & Pairs & Eval. months & MSE & Gain & $p_{\mathrm{shift}}$ \\
\midrule
Count $P^{\mathrm{c}}$
& Baseline $[P^{\mathrm{c}}_{t-1}]$
& 22 & 12 & $0.0033$ & -- & -- \\

Count $P^{\mathrm{c}}$
& $[P^{\mathrm{c}}_{t-1},x^{\mathrm{any}}_t]$
& 22 & 12 & $0.0009$ & $+74.2\%$ & $0.045$ \\

Count $P^{\mathrm{c}}$
& Feature selected within fold
& 22 & 12 & $0.0009$ & $+73.4\%$ & $0.091$ \\

\midrule

Weighted $P^{\mathrm{w}}$
& Baseline $[P^{\mathrm{w}}_{t-1}]$
& 22 & 12 & $0.0094$ & -- & -- \\

Weighted $P^{\mathrm{w}}$
& $[P^{\mathrm{w}}_{t-1},x^{\mathrm{any}}_t]$
& 22 & 12 & $0.0071$ & $+23.8\%$ & $0.045$ \\

Weighted $P^{\mathrm{w}}$
& Feature selected within fold
& 22 & 12 & $0.0090$ & $+4.0\%$ & $0.55$ \\
\bottomrule
\end{tabular}
\caption{Rolling-origin contemporaneous nowcasting. The within-fold rows re-select the language feature using only the training observations available at each evaluation step.}
\label{tab:nowcast_appendix}
\end{table*}

For the count index, adding current-month scalar prevalence reduces out-of-sample error by $74.2\%$ and exceeds all circular shifts of the feature sequence. The augmented model improves 9 of the 12 evaluation months. No single month contributes more than 45\% of the total reduction in squared error, and removing any one evaluation month leaves the gain between approximately $64\%$ and $80\%$.

Because scalar prevalence was identified after inspecting the full association analysis, we repeat the nowcast while selecting the language feature separately within each training fold. At each evaluation step, the feature with the largest absolute training-set correlation with the target is selected from the nine-feature family. For the count index, this procedure selects either scalar prevalence or the share above $0.5$ at every step and recovers a similar gain of $73.4\%$.

The weighted result is weaker. The fixed scalar-prevalence model reduces error by $23.8\%$, but the improvement is more concentrated across months. Seven of the 12 evaluation months improve, one month accounts for a large share of the total gain, and leave-one-month-out gains range from approximately $9\%$ to $43\%$. When the language feature is selected within each training fold, the gain falls to $4.0\%$.

June 2019 lies within the burn-in period and is therefore used for training but is never an evaluation month in the rolling-origin analysis.

The count-index nowcast should nevertheless be interpreted cautiously. The baseline MSE is small in absolute terms, the evaluation contains only 12 held-out months, and the count index itself has a borderline dependence on mean degree. The result therefore supports a contemporaneous language--structure relationship rather than a claim of temporal precedence or causality.
\end{document}